\documentclass[preprint2]{aastex}

\usepackage{graphicx}

\slugcomment{}

\shorttitle{Optical spectroscopy of AGN selected by {\em Spitzer} colors}
\shortauthors{Lacy et al.}

\begin{document}


\title{Optical spectroscopy and X-ray detections of
a sample of quasars and AGN selected in the mid-infrared from 
two {\em Spitzer} wide-area surveys.}


\author{M.\ Lacy\altaffilmark{1}, 
A. Petric\altaffilmark{1,3},
A.\ Sajina\altaffilmark{1},  
G.\ Canalizo\altaffilmark{2}, 
L.J.\ Storrie-Lombardi \altaffilmark{1},
L.\ Armus\altaffilmark{1}, D. Fadda\altaffilmark{1},
F.R.\ Marleau\altaffilmark{1}} 
\altaffiltext{1}{Spitzer Science Center, Caltech, Mail Code 220-6,
Pasadena, CA 91125; mlacy@ipac.caltech.edu,sajina@ipac.caltech.edu, 
lisa@ipac.caltech.edu}
\altaffiltext{2}{University of California, Riverside}
\altaffiltext{3}{Columbia University of New York}

\begin{abstract}
We present optical spectroscopy of a sample of 77 luminous 
AGN and quasars selected
on the basis of their mid-infrared colors. Our objects are selected from the
{\em Spitzer} Extragalactic 
First Look Survey and SWIRE XMM-LSS fields, with a typical
24$\mu$m flux density of 5mJy. The median redshift is 0.6, with a range of 
$\sim 0.05-4$. Only 33\% (25/77) of these objects
are normal type-1 quasars, with no obscuration. 44\% (34/77) are type-2 
objects, with high-ionization, narrow emission lines, 14\% (11/77) 
are dust-reddened 
type-1 quasars, showing broad lines but a dust-reddened or unusually
weak quasar 
continuum. 9\% (7/77) show no sign of an AGN in the optical spectrum, having
either starburst spectra 
or spectra which could be of either starburst or LINER types. 
These latter objects are analogous to the 
X-ray detected population of AGN with weak or non-existent optical 
AGN emission (the ``XBONGs'').
21 of our objects from the SWIRE field fall within moderately-deep
XMM exposures. All the unobscured quasars, and about half the obscured quasars
are detected in these exposures. 
This sample, when taken together with other samples of 
{\em Spitzer} selected AGN and quasars, and results from X-ray studies, 
confirms that obscured AGN dominate the AGN and quasar number counts of 
all rapidly-accreting supermassive black hole systems, at least for 
$z\stackrel{<}{_{\sim}}4$. This implies a high radiative efficiency for the
black hole accretion process. 

\end{abstract}



\section{Introduction}

Quasars whose optical emission is hidden by dust in the optical and gas in the
soft X-ray have always been hard to find. However, a large type-2 (hidden) 
quasar population has long been
predicted, both from the statistics of Seyfert-1 and 2 galaxies, and from 
those of radio galaxies and radio-loud quasars. A large population 
of hidden AGN is also predicted from models of the X-ray background 
(e.g.\ Comastri et al.\ 1995; Worsley et al.\ 2005), though most models
predict these AGN to be of Seyfert rather than quasar luminosity.
Some success in finding dust obscured quasars has been obtained using 
selection based on the Two Micron All Sky Survey (2MASS)
to find lightly dust-reddened (rest frame extinctions $A_V \sim 2$) 
quasars, based either on 2MASS colors alone (Cutri et al.\ 2001), or on
2MASS combined
with a radio detection and very red optical-IR colors (Glikman et al.\ 2005). 
Only recently, however, have large 
samples of type-2 quasars, with rest-frame $A_V\stackrel{>}{_{\sim}}5$ 
been found. Zakamsaka et al.\ (2003) used the 
Sloan Digital Sky Survey (SDSS) to find objects with the high-ionization 
narrow-line spectra characteristic of type-2 AGN.  
Combining {\em ISO} and 2MASS data allowed Haas et al.\ (2004a) and Leipski et 
al.\ (2005) to find significant numbers of obscured AGN, including several 
type-2 quasars. {\em Spitzer} data has been
used to find highly-obscured AGN, both through mid-infrared 
color selection (Lacy et al.\ 2004 [hereafter Paper I], 2005a;
Stern et al.\ 2005; Alonso-Herrero et al.\ 2005; Polletta et al.\ 2006) and 
joint radio-infrared selection
(Mart\'\i nez-Sansigre et al.\ 2005, 2006; Weedman et al.\ 2006; Donley et al.\ 2005). 
Type-1 quasars selected through joint Spitzer and optical 
colors and/or morphologies have been targeted by several groups 
(Brown et al.\ 2005; Hatziminaoglou et al.\ 2005; Siana et al.\ 2006). 
Also, the 
mid-infrared properties of X-ray selected AGN have been discussed by 
Treister et al.\ (2005) and Barmby et al.\ (2005). 

Mid-infrared selection of AGN is effective because dust heated by the AGN
forms a ``calorimeter'', whose luminosity is not as strongly
affected by obscuration as that in most other wavelength bands (except the
radio) (e.g.\ Spinoglio et al.\ 1995; Meisenheimer et al.\ 2001; Ogle et al.\ 2006). 
Richards et al.\ (2006) show that the bolometric correction in the
mid-infrared varies little for different subsamples of type-1
quasars (e.g.\ optically or infrared 
luminous, optically-red or optically blue), or luminosity, 
and that about 15\% of the 
bolometric luminosity of a quasar is emitted at mid-infrared wavelengths.
Integration over the Richards et al.\ (2006) mean quasar SED as a whole
suggests that about 40\% of the emission from a typical quasar is 
reprocessed by dust in the infrared. 
This means that mid-infrared selection of AGN is probably a good proxy for 
selection on the basis of bolometric luminosity, at least for those
objects which are not so highly obscured that their mid-infrared luminosity 
is significantly supressed. Much of 
the dust emission is thought to arise in the ``obscuring torus'' required
by Unified Schemes (e.g.\ Antonucci 1993), though it is becoming 
increasingly clear that host dust in the narrow line region of AGN
also contributes significantly (e.g.\ Mason et al.\ 2006). 
AGN are an important global contributor to the mid-infrared luminosity
density.
Brand et al.\ (2006) show that the luminosity density of mid-IR emission 
in samples of MIPS 24$\mu$m sources selected at mJy levels 
is dominated by AGN, only at submJy
levels do star forming galaxies begin to dominate. 

In this paper, we focus on optical spectroscopic follow-up of AGN candidates
selected according to the mid-infrared 
color criteria presented in Paper I. By selecting
relatively bright objects from wide-area 
{\em Spitzer} surveys we are able to find many obscured 
AGN with mid-infrared luminosities typical of 
quasars at moderate redshifts. 
These objects provide a good complement 
to the mostly lower-luminosity or higher redshift 
AGN found in surveys such as the Great 
Observatories Origins Deep Survey (GOODS) and the Chandra Deep fields.
Preliminary results from the spectroscopic follow-up of the sample presented
in Paper 1 were given in Lacy et al.\ (2005a). In this paper, we describe
the selection of samples of mid-infrared selected 
AGN candidates flux-limited at 
24$\mu$m from both the Extragalactic
First Look Survey (XFLS) and the {\em Spitzer} Wide-area Infrared 
Extragalactic Survey (SWIRE) XMM-LSS field (Lonsdale et al.\ 2005), 
and their spectrosopic follow-up observations.   
We classify the optical spectra according to the type of 
AGN, and use them to place a lower bound on the fraction of 
obscured quasars
and luminous Seyfert galaxies. We also investigate the X-ray fluxes
of those objects which fall on archival observations made with the 
X-ray Multiple Mirror telescope ({\em XMM-Newton}). 
A more detailed analysis of the Spectral Energy Distributions (SEDs) 
and X-ray properties of the AGN will be carried
out in a later paper (Petric et al.\ 2006).

We assume a cosmology with $\Omega_{\rm M}=0.3$, $\Omega_{\Lambda}=0.7$
and $H_0=70 {\rm kms^{-1}Mpc^{-1}}$.

\section{The Spitzer XFLS and SWIRE AGN samples}

We selected samples of AGN by mid-infrared color from the XFLS and 
SWIRE XMM fields. The basis of our 
color selection technique is described in Sajina, Lacy \& Scott (2005) 
and Paper I, though the samples in this paper are flux-limited at 24$\mu$m
rather than at 8$\mu$m as was the case for the sample of Paper 1. 
To obtain the XFLS sample, we began by 
matching the XFLS 24$\mu$m catalog of Fadda et al.\ (2006) to the 
Infrared Array Camera (IRAC) 4-band catalog of Lacy et al.\ (2005). All
MIPS 24$\mu$m detections with detections in all four IRAC bands were
then plotted on the color-color diagram of Figure 1. For the SWIRE XMM sample,
we used the bandmerged SWIRE IRAC/MIPS 24$\mu$m catalog available
from the {\em Spitzer} popular products website.
The color cut used to select AGN candidates 
was ${\rm lg}(S_{5.8}/S_{3.6}) > -0.1$ and ${\rm lg}(S_{8.0}/S_{4.5} > -0.2$
and ${\rm lg}(S_{8.0}/S_{4.5}) \leq 0.8  {\rm lg}(S_{5.8}/S_{3.6}) + 0.5$.
The ${\rm lg}(S_{5.8}/S_{3.6})$ cut was moved by $+0.1$ compared to Paper I.
The exact position of this cut is not critical, and it was felt that 
shifting this would help remove non-AGN contaminants, particularly as the
final 5.8$\mu$m fluxes in the XFLS catalogue were increased by $\approx 10$\% 
in the final version of the XFLS catalog relative to the catalog used in 
Paper I. We discuss the exact 
placement of the color cuts with respect to the completeness of the AGN 
selection in section 5.3. Candidates were selected down to limiting 24$\mu$m
fluxes of 4.4mJy in the XFLS and 6.6mJy in the SWIRE XMM field.

The 24$\mu$m flux limited sample should be an improvement on 8$\mu$m selection
by making us 
more sensitive to highly-reddened objects, and high
redshift objects, where the strong $k$-correction on the mid-infrared
dust emission means that they drop out of flux-limited samples selected at 
shorter wavelengths. 
The disadvantage of this selection is that we will
be missing objects at $z\sim 1.5$ which have strong silicate absorption in the 
24$\mu$m band, however, as our median redshift is $\approx 0.6$ only a 
few objects will be missed due to this effect. The XFLS sample is complete
for objects falling in the IRAC ``AGN wedge'' color selection to a flux limit of 4.6mJy at 
$24\mu$m, and the SWIRE sample to a flux limit of 6.6mJy (note that 
all MIPS sources down to these flux limits have 4-band IRAC detections). 
Approximately 20\% of MIPS 24$\mu$m sources at these flux levels fall into
our AGN-dominated part of the color-color plot.

A few stars and nearby galaxies have been removed from the samples. Most stars
were only present in the initial selection due to photometric errors
related to saturation in the shorter IRAC bands, though one star
(SSTXFLS J172432.8+592646, see Paper I) 
has the correct colors to fall in the sample, 
presumably due to a dust disk. Very low redshift 
($z\stackrel{<}{_{\sim}} 0.05$) galaxies are also 
possible contaminants due to the presence of 6.2$\mu$m
PAH emission in the 5.8$\mu$m
band of IRAC (Sajina, Lacy \& Scott 2005). 
The interacting galaxy Arp 54 appeared in the selection of 
the SWIRE AGN sample, but has been removed for this reason.

\begin{table*}
{\scriptsize
\caption{Observing log for the XFLS AGN sample}
\begin{tabular}{lccclcc}
Name & Date observed & Telescope/Instrument&Exposure& Slit &Airmass &PA \\ \hline
SSTXFLS J171115.2+594906&2004 August 19&Palomar 200/DbleSpec&$2\times 600$&2.0&1.32&105\\
SSTXFLS J171147.4+585839&2004 August 18&Palomar 200/DbleSpec&$2\times 600$&2.0&1.44&95\\
SSTXFLS J171233.4+583610&2005 Oct 30&Palomar 200/Cosmic&2$\times$900&2.0&1.40&90\\
SSTXFLS J171302.3+593610&2005 June 01& Palomar 200/Cosmic&2$\times$ 1200&1.5&1.15&0\\
SSTXFLS J171313.9+603146&2004 July 18&Keck II/ESI&1$\times$600&1.0&1.55&127\\
                        &            &                 &              &   & &\\
SSTXLFS J171324.2+585549&2004 August 19&Palomar 200/DbleSpec&$2\times 600$&2.0&1.35&100\\
SSTXFLS J171325.1+590531&2005 Nov 01&Palomar 200/Cosmic&1$\times$ 900&1.5&1$\times$90076\\
SSTXFLS J171331.5+585804&2005 June 02&Palomar 200/Cosmic&2$\times$1200&1.5&1.15&165\\
SSTXFLS J171335.1+584756&2005 June 01&Palomar 200/Cosmic&2$\times$600&1.5&&90\\
SSTXFLS J171345.5+600730&2005 Oct 30&Palomar 200/Cosmic&2$\times$1800&2.0&1.70&90\\
                        &            &                 &              &   & &\\
SSTXFLS J171345.5+600730&2006 June 27&Palomar 200/Cosmic&3$\times$1800&2.0&1.13&180\\
SSTXFLS J171419.9+602724&2005 June 02&Palomar 200/Cosmic&1$\times$1200&1.5&1.18&45\\
SSTXFLS J171430.7+584225&2005 Oct 31&Palomar 200/Cosmic&2$\times$900&1.5&1.86&190\\
SSTXLFS J171513.8+594638&2002 June 15&Shane/Kast &1$\times$ 1200&1.5&1.24&73\\
SSTXFLS J171530.7+600216&2005 June 01& Palomar 200/Cosmic&1$\times$600&1.5&1.15&140\\S
                        &            &                 &              &   & &\\
SSTXFLS J171650.6+595752&2005 June 02 &Palomar 200/Cosmic&1$\times$600 &1.5&1.23&110\\
SSTXFLS J171708.6+591341&2005 June 02&Palomar 200/Cosmic&1$\times$900 &1.5&1.26&110\\
SSTXFLS J171750.7+584745&2005 Oct 30&Palomar 200/Cosmic&2$\times$1800&2.0&1.50&90\\
SSTXFLS J171754.6+600913&2005 June 01&Palomar 200/Cosmic&2$\times$ 1200&1.5&1.12&0\\ 
SSTXFLS J171831.5+595317&2004 July 13&WHT/ISIS&2$\times$600&2.0&1.19&5\\
                        &            &                 &              &   & &\\
SSTXFLS J171839.6+593359&2002 June 11&Shane/Kast &1$\times$ 1200&1.5&1.26&109\\
SSTXFLS J171913.5+584508&2005 June 01&Palomar 200/Cosmic&2$\times$600&1.5&1.18&140\\
SSTXFLS J172044.8+582923&2005 June 02&Palomar 200/Cosmic&2$\times$1800&2.0&1.11&90\\
SSTXFLS J172050.4+591511&2004 July 13&WHT/ISIS&2$\times$900&2.0&1.16&85\\
SSTXFLS J172050.4+591511&2006 June 28&Palomar 200/Cosmic&3$\times$1800&1.5&1.15&210\\
                        &            &                 &              &   & &\\
SSTXFLS J172123.1+601214&2004 July 18&Keck II/ESI&1$\times$600&1.0&1.57&125\\
SSTXFLS J172219.5+594506&2005 June 02&Palomar 200/Cosmic&2$\times$600&1.5&1.48&90\\
SSTXFLS J172228.1+601526&2005 June 02&Palomar 200/Cosmic&1$\times$1200&1.5&1.24&45\\
SSTXFLS J172245.0+590328&2005 Nov 01 & Palomar 200/Cosmic&2$\times$1200&1.5&1.55&100\\
SSTXFLS J172248.9+583256&2005 June 02&Palomar 200/Cosmic&2$\times$1200&1.5&1.12&45\\
                        &            &                 &              &   & &\\
SSTXFLS J172248.9+583256&2006 May 26&IRTF/Spex&22$\times$120&0.5&1.48&100\\
SSTXFLS J172328.4+592947&2004 August 19&Palomar 200/DbleSpec&$2\times 600$&2.0&1.5&90\\
SSTXFLS J172542.3+595317&2005 June 02&Palomar 200/Cosmic&2$\times$ 1200&1.5&1.34&90\\

\end{tabular}
}
\end{table*}

\begin{table*}
{\scriptsize
\caption{Observing log for the SWIRE XMM-LSS sample}
\begin{tabular}{lccclcc}
Name & Date observed & Telescope/Instrument&Exposure& Slit &Airmass &PA \\ \hline
SWIRE2 J021638.21-042250.8&2005 Oct 30&Palomar 200/Cosmic&1$\times$ 900&1.5&1.45&125\\  
SWIRE2 J021657.77-032459.7&2005 Oct 29&Palomar 200/Cosmic&1$\times$ 600&1.5&1.84&225\\ 
SWIRE2 J021729.06-041937.8&2005 Oct 30&Palomar 200/Cosmic&2$\times$1800&1.5&1.4&20\\   
SWIRE2 J021749.00-052306.9&2005 Nov 01&Palomar 200/Cosmic&2$\times$1800&1.5&1.30&7\\  
SWIRE2 J021759.91-052056.1&2005 Oct 31&Palomar 200/Cosmic&2$\times$1200&1.5&1.28&20\\  
                        &            &                 &              &   & &\\
SWIRE2 J021809.45-045945.9&2005 Oct 31&Palomar 200/Cosmic&2$\times$1200&1.5&1.31&20\\  
SWIRE2 J021859.74-040237.2&2005 Oct 30&Palomar 200/Cosmic&1$\times$ 600&1.5&1.89&312\\
SWIRE2 J021909.60-052512.9&2005 Oct 29&Palomar 200/Cosmic&1$\times$ 600&1.5&1.31&150\\ 
SWIRE2 J021938.70-032508.2&2005 Nov 01&Palomar 200/Cosmic&1$\times$600&1.5&1.40&42\\ 
SWIRE2 J021939.08-051133.8&2005 Oct 29&Palomar 200/Cosmic&1$\times$ 600&1.5&1.36&90\\ 
                        &            &                 &              &   & &\\
SWIRE2 J022005.93-031545.7&2005 Nov 01&Palomar 200/Cosmic&1$\times$600&1.5&1.55&47\\ 
SWIRE2 J022012.21-034111.8&2005 Nov 01&Palomar 200/Cosmic&1$\times$600&1.5&1.74&47\\ 
SWIRE2 J022039.48-030820.3&2005 Oct 31&Palomar 200/Cosmic&1$\times$ 900&1.5&1.50&140\\   
SWIRE2 J022133.82-054842.8&2005 Nov 01&Palomar 200/Cosmic&1$\times$600&1.5&1.75&145\\ 
SWIRE2 J022211.57-051308.1&2005 Nov 01&Palomar 200/Cosmic&2$\times$1800&1.5&1.31&341\\  
                        &            &                 &              &   & &\\
SWIRE2 J022225.86-050015.1&2005 Oct 31&Palomar 200/Cosmic&1$\times$600&1.5&2.24&48\\  
SWIRE2 J022255.87-051351.7&2005 Oct 31&Palomar 200/Cosmic&1$\times$900&1.5&1.49&30\\  
SWIRE2 J022301.97-052335.8&2005 Nov 01&Palomar 200/Cosmic&2$\times$1200&1.5&1.46&341\\  
SWIRE2 J022306.74-050529.1&2005 Oct 30&Palomar 200/Cosmic&2$\times$ 900&1.5&1.76&315\\ 
SWIRE2 J022310.42-055102.7&2005 Oct 29&Palomar 200/Cosmic&2$\times$ 1800&2.0&1.24&90\\ 
                        &            &                 &              &   & &\\
SWIRE2 J022348.96-025426.0&2005 Nov 01&Palomar 200/Cosmic&1$\times$900&1.5&1.47&145\\ 
SWIRE2 J022356.49-025431.1&2005 Oct 29&Palomar 200/Cosmic&2$\times$1800&2.0&1.24&90\\   
SWIRE2 J022422.27-031054.7&2005 Oct 31&Palomar 200/Cosmic&1$\times$600&1.5&1.67&48\\  
SWIRE2 J022431.58-052818.8&2005 Oct 31&Palomar 200/Cosmic&1$\times$ 600&1.5&1.7&310\\   
SWIRE2 J022438.97-042706.3&2005 Nov 01&Palomar 200/Cosmic&1$\times$600&1.5&1.95&45\\  
                        &            &                 &              &   & &\\
SWIRE2 J022455.77-031153.6&2005 Oct 31&Palomar 200/Cosmic&1$\times$900&1.5&1.52&40\\ 
SWIRE2 J022508.33-053917.7&2005 Oct 31&Palomar 200/Cosmic&1$\times$ 900&1.5&1.32&26\\   
SWIRE2 J022542.02-042441.1&2005 Oct 30&Palomar 200/Cosmic&2$\times$ 900&1.5&1.6&40\\   
SWIRE2 J022600.01-035954.5&2005 Oct 31&Palomar 200/Cosmic&1$\times$ 900&1.5&1.45&125\\   
SWIRE2 J022612.67-040319.4&2005 Oct 31&Palomar 200/Cosmic&1$\times$600&1.5&1.88&48\\ 
                        &            &                 &              &   & &\\
SWIRE2 J022700.77-042020.6&2005 Oct 29&Palomar 200/Cosmic&1$\times$ 600&1.5&1.32&150\\
\end{tabular}
}
\end{table*}

Tables 1 and 2 contain the observing logs for our samples. Most data 
were taken with the Cosmic instrument on the Palomar 200-inch. 
Some spectra were obtained with the ISIS instrument 
at the 4.2m William Herschel Telescope 
(WHT) on La Palma, with ESI at the Keck II telescope, or
with the Kast spectrograph at the 3m Shane 
Telescope at Lick Observatory. SSTXFLS J172248.9+583256, for which we 
failed to obtain a redshift in the optical was observed in the near-infrared
using the low dispersion mode of the Spex instrument at the 
Infrared Telescope Facility (IRTF) (Rayner et al.\ 2003). Data were
analysed in IRAF using standard routines for flat fielding, wavelength
calibration and sky subtraction. No attempt has been made
to correct for reddening. An extinction correction has been applied based
on airmass, but no attempt has been made to correct the telluric 
absorption features. The spectra are approximately photometrically
calibrated, but variable seeing and cirrus cloud means that we expect our 
spectrophotometric accuracy to be no better than 50\%. In the XFLS field, most
quasars have spectra from the
Sloan Digital Sky Survey (SDSS), which have been used where
they exist. A few of the quasars in the SWIRE field
have redshifts from the literature.
Several of our XFLS sources have spectra in the Multiple Mirror Telescope 
(MMT) spectroscopic survey of Papovich et al\ (2006; hereafter P06), 
including one object
missed from our original selection but which is within the flux-limited
sample (SSTXFLS J171454.4+584948). The spectra of all the objects with the
exception of the SWIRE quasars with redshifts from the literature
are shown in Figures 2 and 3. Objects with spectral classifications 
from both this paper and Papovich et al.\ (2006) are discussed in Appendix
A. For completeness, Appendix B contains 
details of the results of spectroscopy on the objects in the 
8$\mu$m sample of Paper I which are too faint at 24$\mu$m to fall in the 
XFLS sample.

\begin{figure}

\plotone{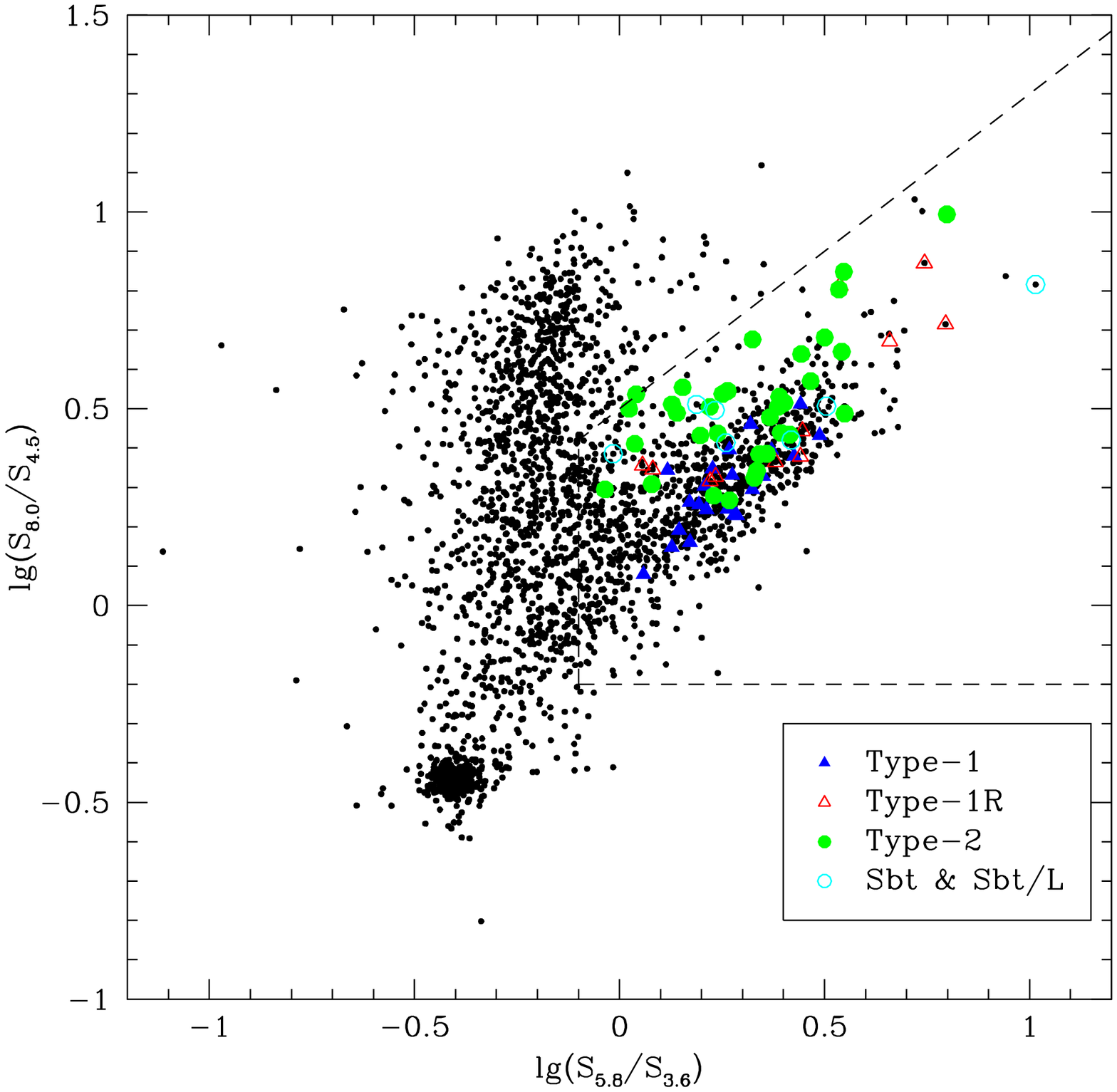}

\caption{IRAC color-color plot for the XFLS field for objects 
detected in MIPS 24$\mu$m and all four IRAC bands. The region 
within the dashed line is that used to select AGN. Objects 
in the AGN samples (both XFLS and SWIRE) are plotted as colored
symbols. Object classifications are discussed in Section 3.
}

\end{figure}


\begin{figure*}

\includegraphics[scale=0.8]{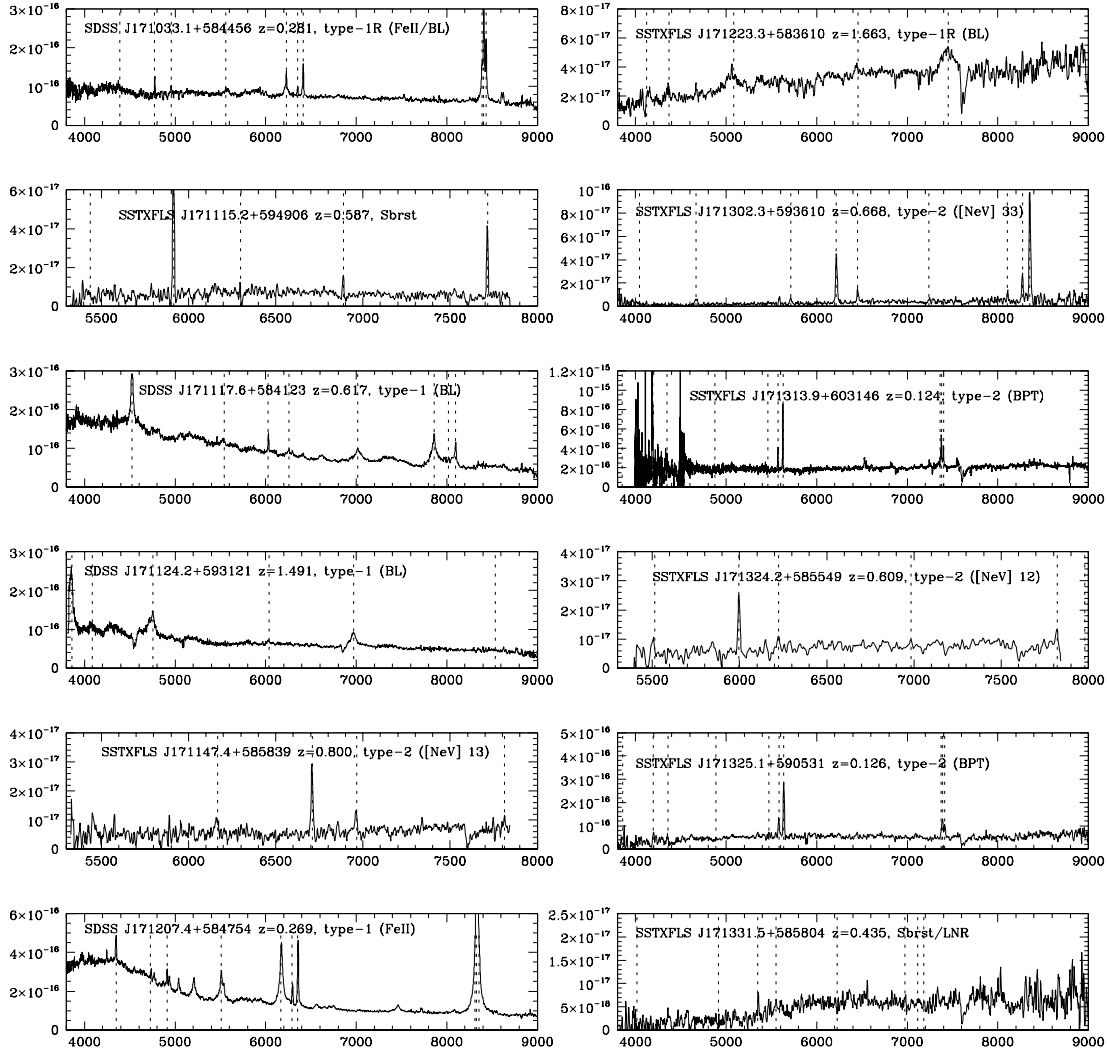}
\caption{Optical spectra of the XFLS AGN candidates. Dotted lines 
indicate the positions of redshifted emission lines (not all of which
are necessarily detected in a given spectrum). The emission 
lines indicated are (from blue to red): Ly$\alpha$, N{\sc v}1240, C{\sc iv}
1549, He{\sc ii}1640, C{\sc iii}]1909, [Ne{\sc iv}]2424, Mg{\sc ii}2798, 
[Ne{\sc v}]3426, [O{\sc ii}]3727, [Ne{\sc iii}]3869, H$\gamma$, H$\beta$,
[O{\sc iii}]4959, [O{\sc iii}]5007, [N{\sc ii}]6548, H$\alpha$ and 
[N{\sc ii}]6584. The spectra are presented in order of Right Ascension. }
\end{figure*}

\setcounter{figure}{1}

\begin{figure*}
\includegraphics[scale=0.8]{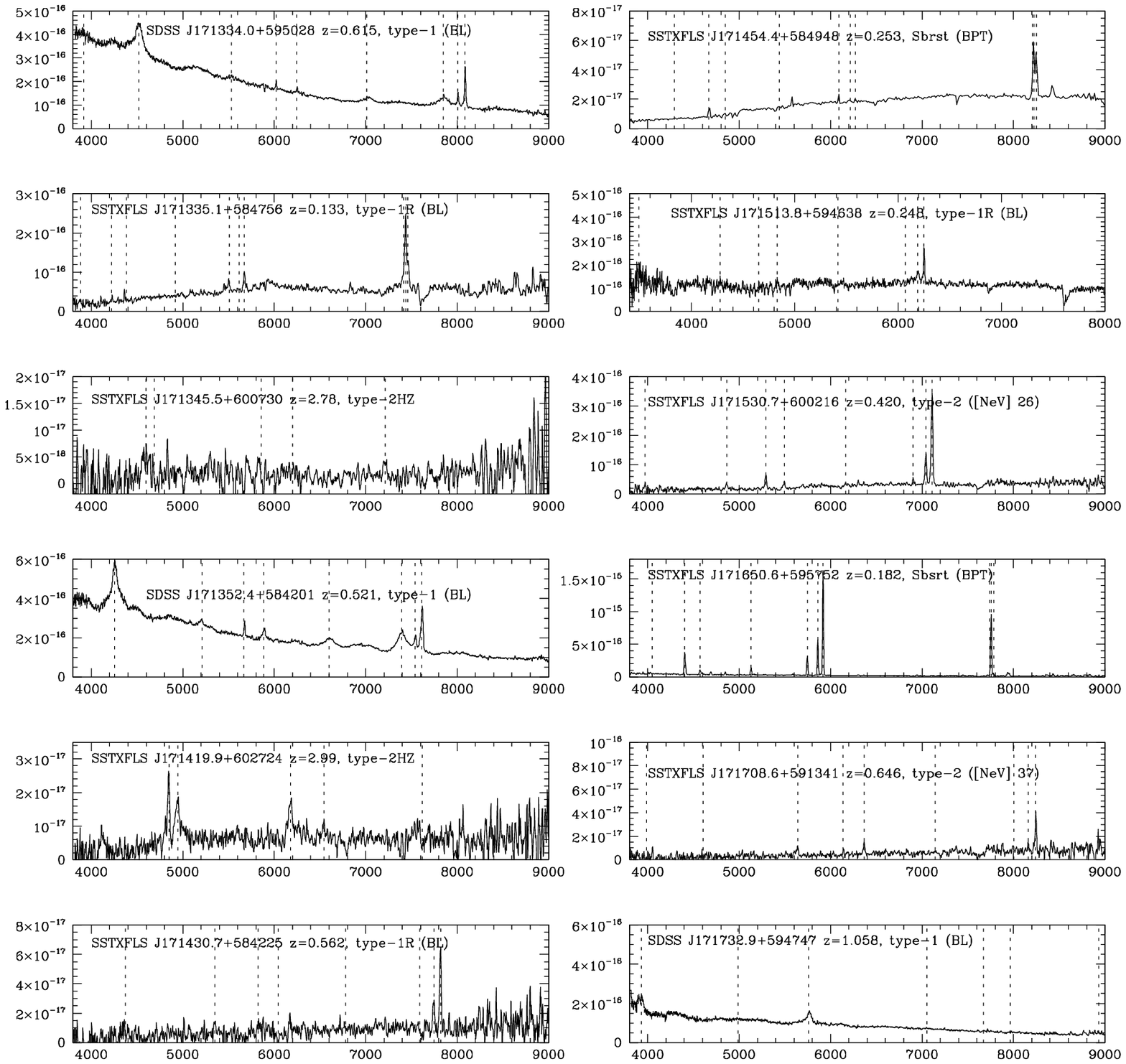}
\caption{Optical spectra of the XFLS AGN candidates, continued}
\end{figure*}

\setcounter{figure}{1}

\begin{figure*}
\includegraphics[scale=0.8]{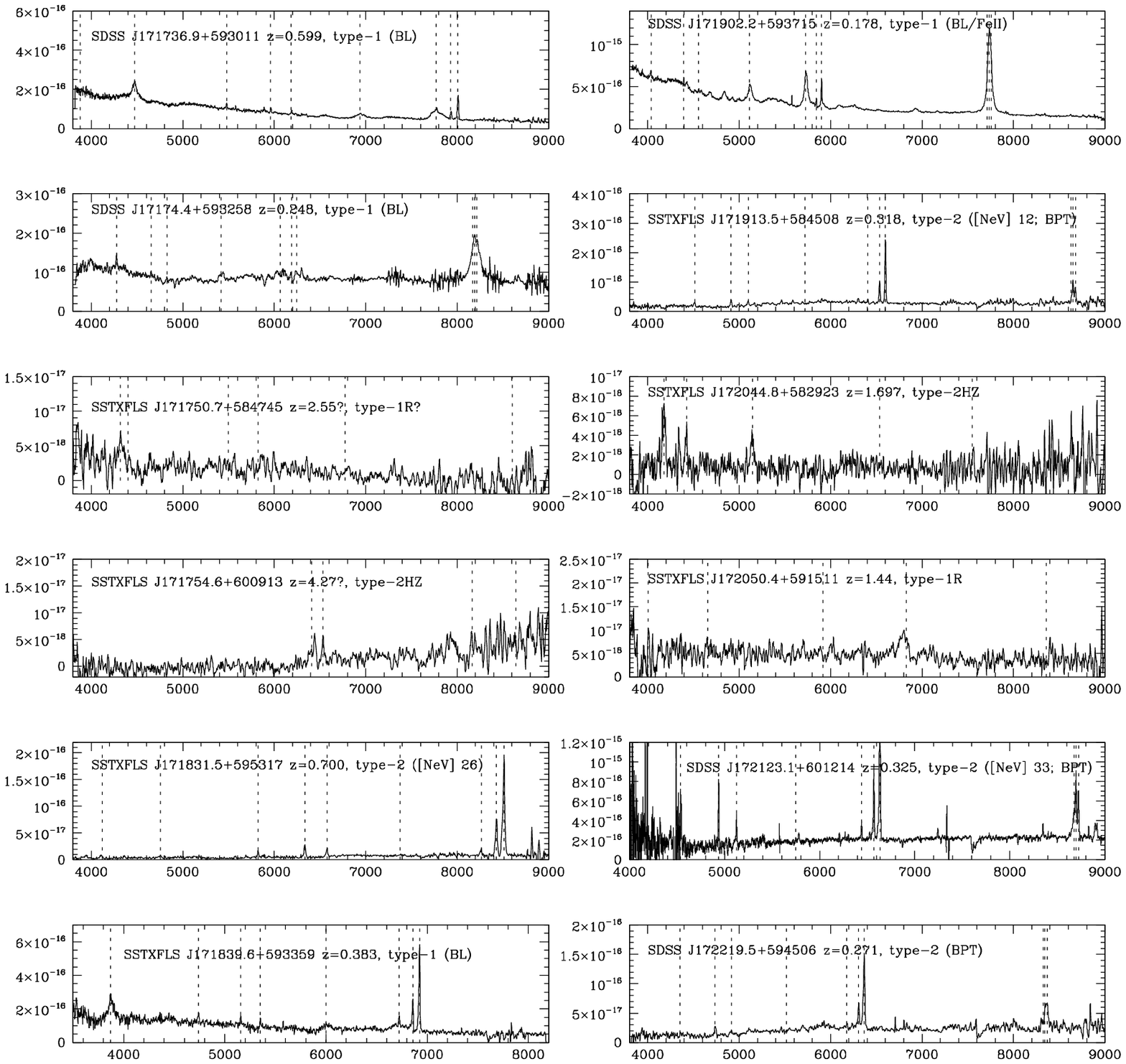}
\caption{Optical spectra of the XFLS AGN candidates, continued}
\end{figure*}

\setcounter{figure}{1}

\begin{figure*}
\includegraphics[scale=0.8]{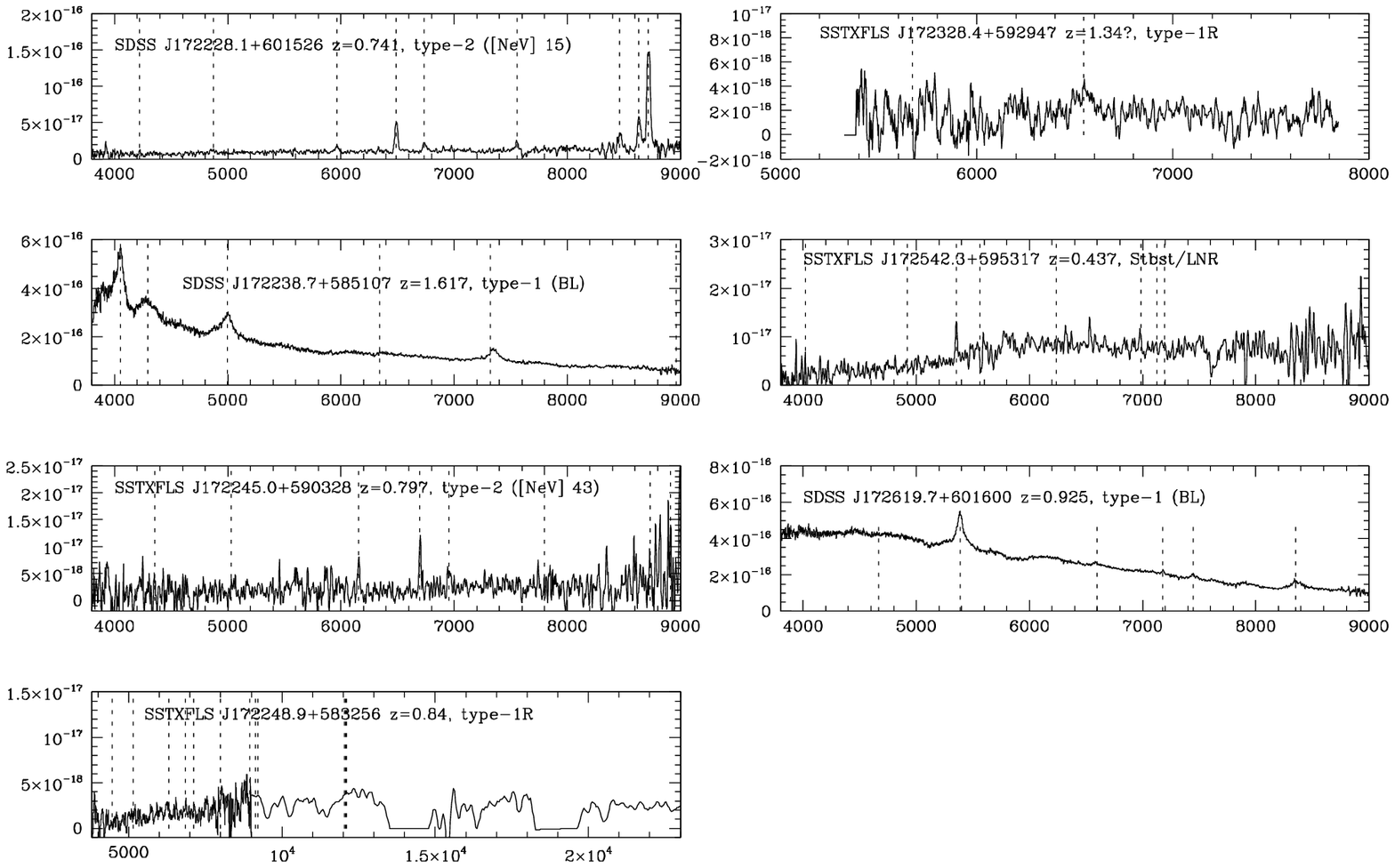}
\caption{Optical spectra of the XFLS AGN candidates, continued}
\end{figure*}


\begin{figure*}
\includegraphics[scale=0.8]{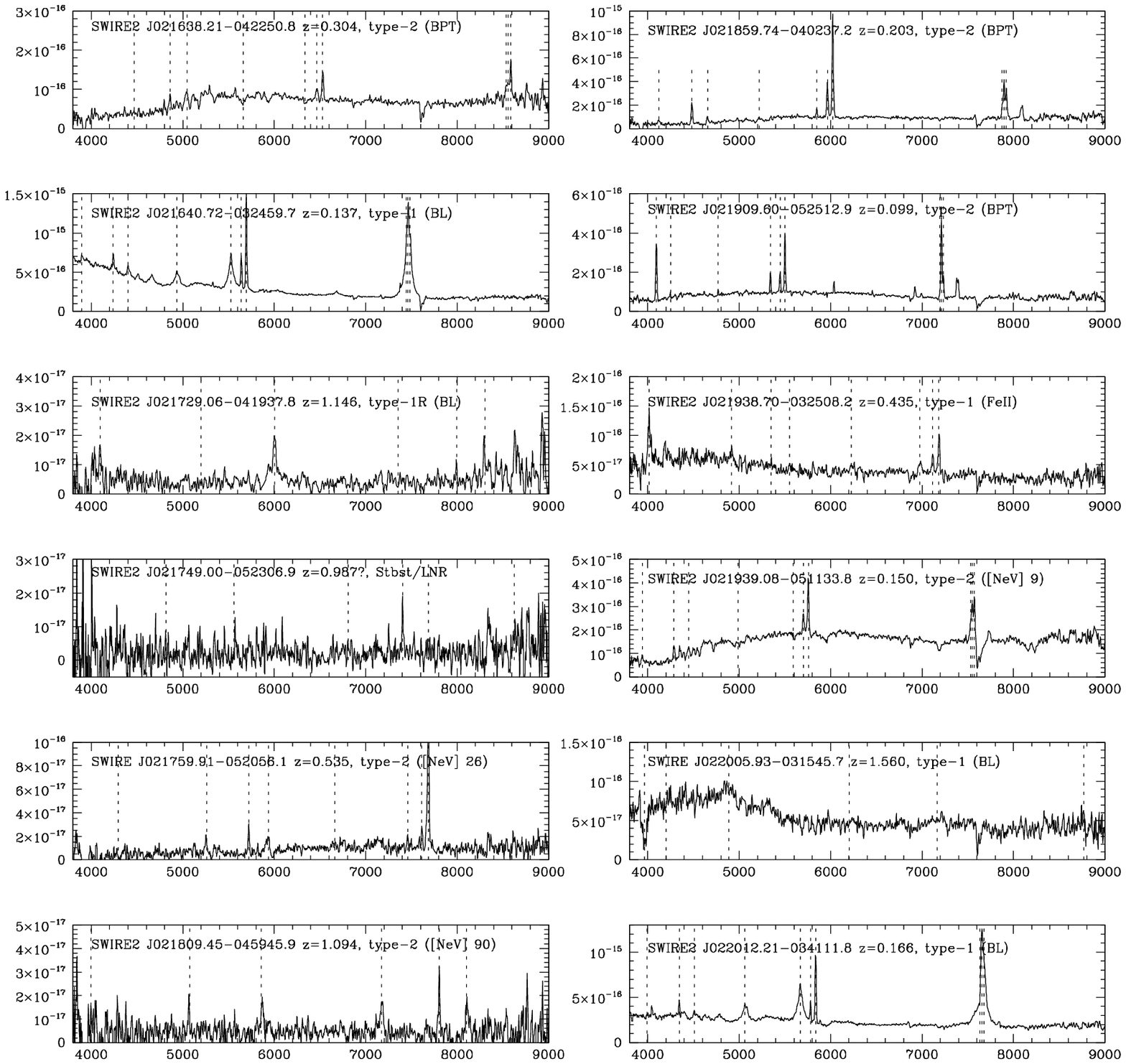}
\caption{Optical spectra of the SWIRE AGN candidates. The position of 
emission lines are indicated as described in the caption to Figure 2.
The spectra are presented in order of Right Ascension.}
\end{figure*}

\setcounter{figure}{2}

\begin{figure*}
\includegraphics[scale=0.8]{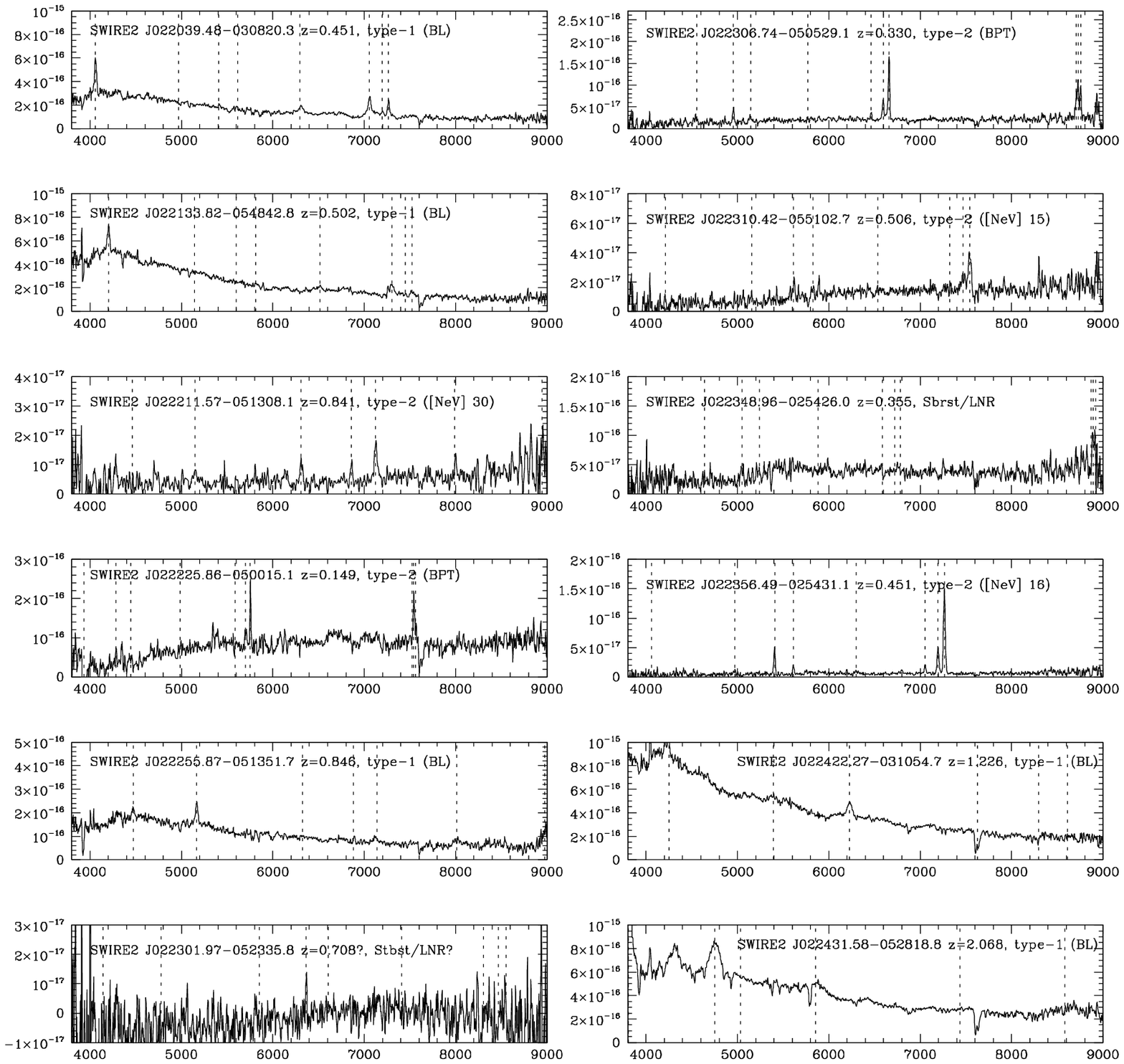}
\caption{Optical spectra of the SWIRE AGN candidates, continued}
\end{figure*}

\setcounter{figure}{2}

\begin{figure*}
\includegraphics[scale=0.8]{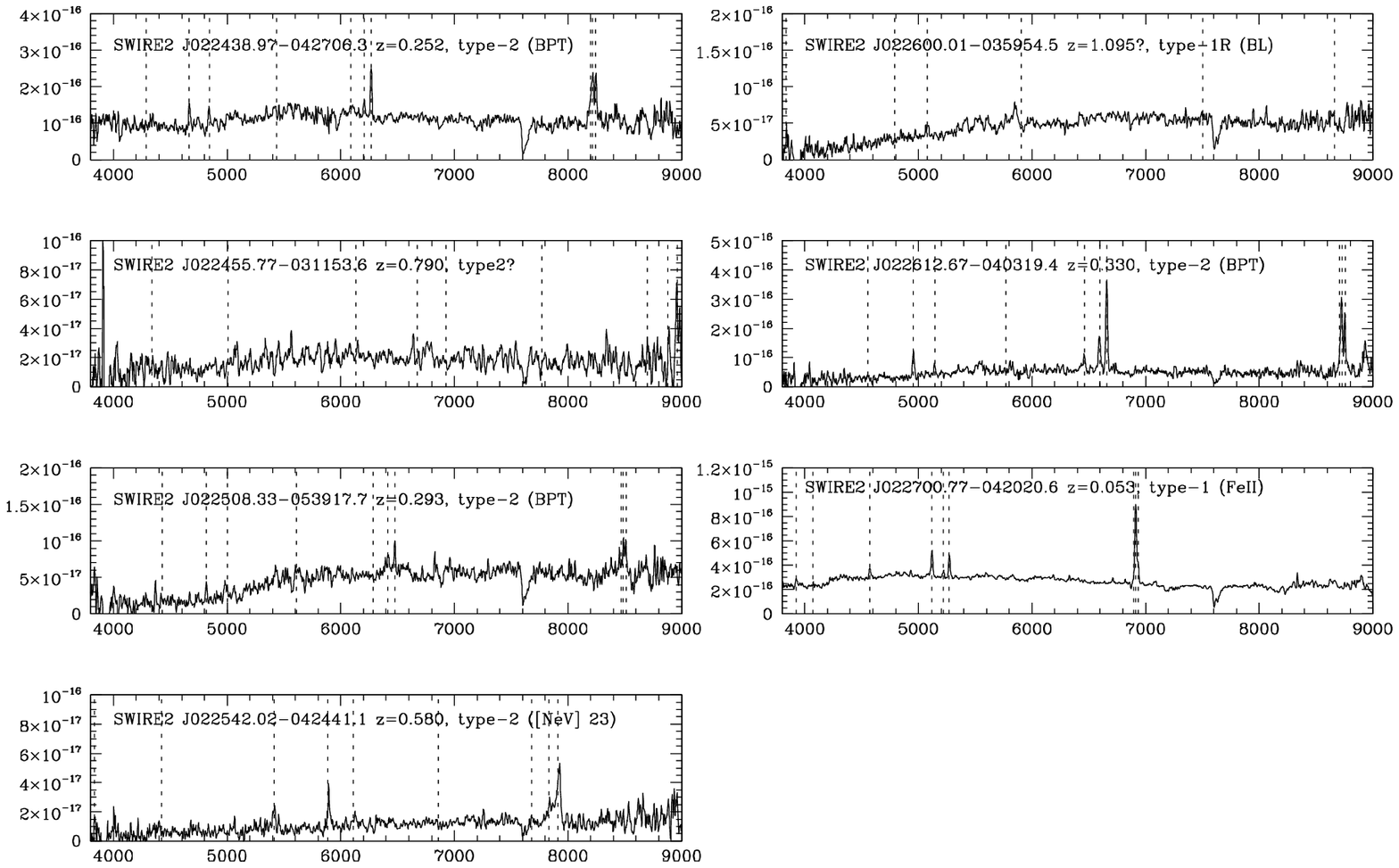}
\caption{Optical spectra of the SWIRE AGN candidates, continued}
\end{figure*}

\begin{figure*}
\centering
\columnwidth=.45\columnwidth
\includegraphics[scale=0.4]{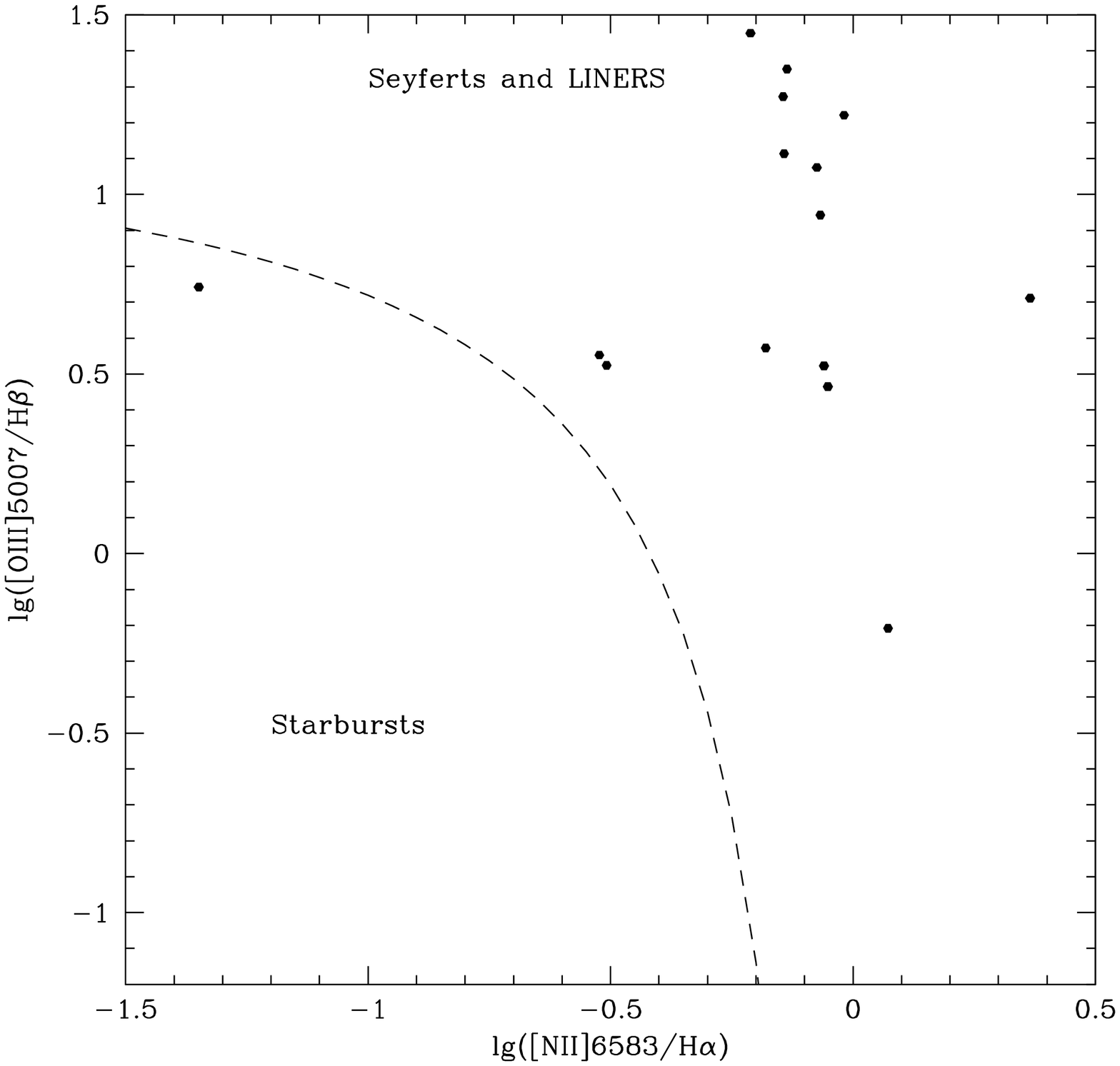}
\hfil
\includegraphics[scale=0.4]{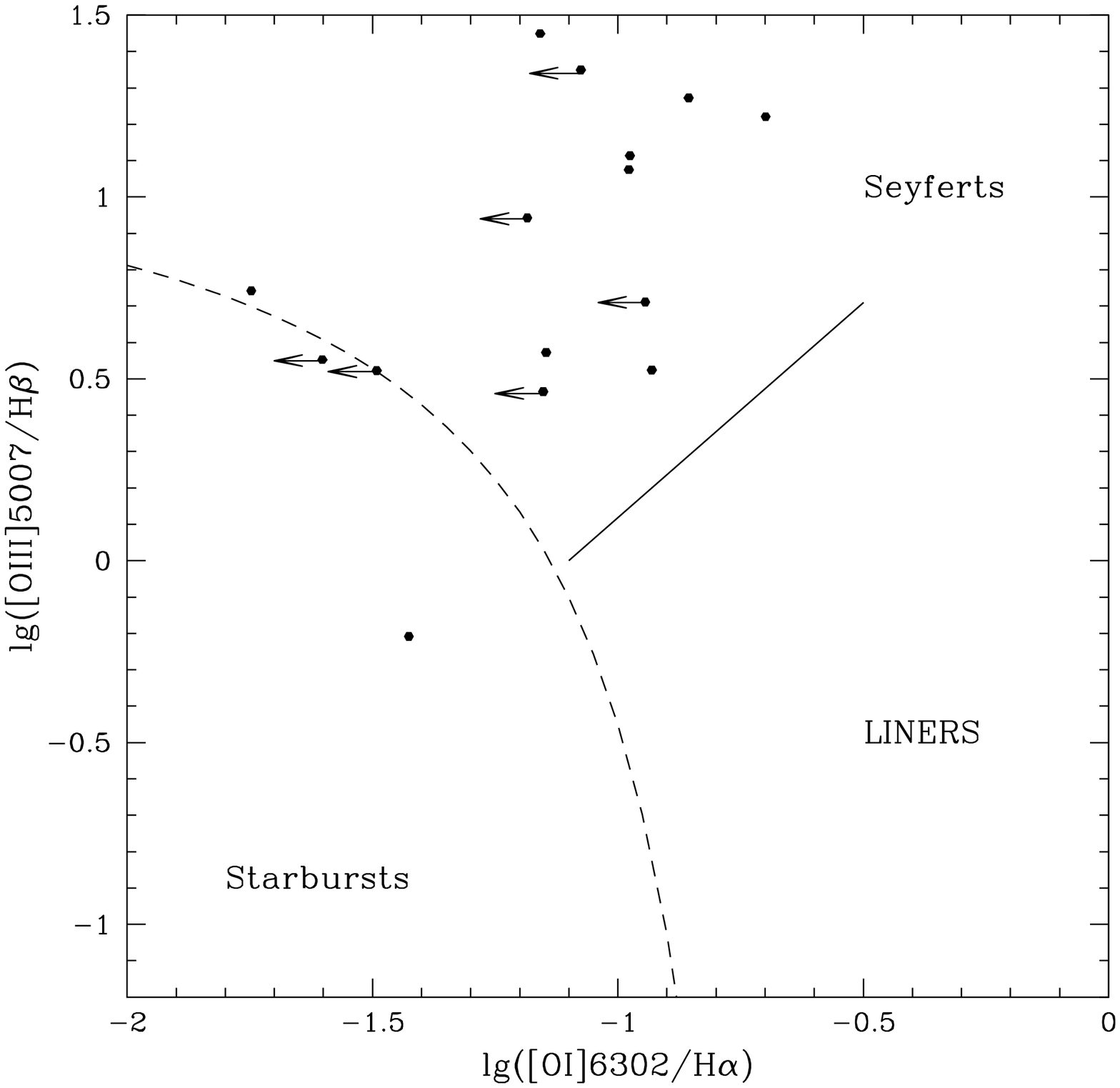}
\caption{Two diagnostic (BPT) diagrams for our lower redshift emission line
galaxies that have H$\alpha$ in the optical window. The dividing lines
are from Kewley et al.\ (2006). Most of our objects plot amongst the AGN, 
there are no LINERs in the sample.}

\end{figure*}

\section{Results}

\subsection{Optical spectral classification of the AGN candidates}

We have used a combination of standard classification schemes to classify our
AGN candidates, principally because the large redshift and luminosity range of our sample
makes any single scheme unworkable. We divide our objects into four main 
classes. Normal quasars/Seyfert-1 galaxies are assigned type 1. The 
obscured quasars are subdivided into three classes. Reddened 
quasars,
showing broad lines in their optical spectra, but with a faint, presumably
dust-reddened quasar continuum and/or clear host galaxy emission 
with no sign of a blue quasar continuum above the host galaxy light 
are classed
type-1R. Objects with high-ionization narrow emission lines which cannot
be produced in a pure starburst scenario are classed as type-2. Objects with
starburst emission line ratios in emission line diagnostic
diagrams (Baldwin, Phillips \& Terlevich 1983; hereafter BPT), and/or low
ionization spectra with strong Balmer lines are classed as starbursts (Sbt). 
Where we have 
a low-ionization spectrum with too few lines
to plot on a diagnostic diagram, or are low ionization and 
have H$\alpha$ beyond the range 
of the optical spectrum they are classed as
starburst/LINER (Sbt/L). Details of the bases of our 
optical classifications are described below. It should be emphasized that these
classes are based on the {\em observed}-frame optical spectra, the rest-frame
optical spectra may give different classifications.
The classifications are given in the final columns of Tables 3 and 4, 
together with their basis. Of our sample of 77 objects, we find 25 objects
with type-1 optical spectra, 11 with type-1R, 
34 with type-2, two with unambiguous starburst spectra and five 
with spectra which could be from either starbursts or LINERs. 

\subsection{Type-1 objects}

Objects with linewidths of at least one permitted emission line 
$>3000$km$^{-1}$ are classified as type-1s. These objects have ``BL'' in 
the classification basis columns in Tables 3 and 4. 
This is sufficient to distinguish 
most of the luminous quasars. However, as our sample spans a wide luminosity
range, several of the lower luminosity, lower redshift type-1 objects have
narrower permitted lines (including a 
couple of candidate narrow-line Seyfert-1
galaxies). We therefore use the presence of 
strong permitted lines (H$\beta$/[O{\sc iii}]5007$> 0.5$) and Fe{\sc ii} 
emission in the 4000-5500\AA$\;$range to classify them as type-1.
These objects have ``Fe{\sc ii}'' in the basis columns.  
The division into normal type-1 or 
reddened, type-1R, quasars is also made.
This is decided on one of 
two criteria. If the optical spectrum is quasar-like, with 
featureless continuum and broad emission lines, but the 
continuum appears significantly 
reddened ($E(B-V) \stackrel{>}{_{\sim}} 0.3$) 
compared to that of a normal quasar (see section 3.5), 
then the object is classified 
at type-1R. Also, some low-$z$ objects were classed as type-1R 
if the optical spectrum contains broad lines, but host galaxy features are 
visible and the object is dominated by host galaxy light in the blue.

\subsection{Type-2 objects}

Objects without broad lines or Fe{\sc ii} emission, but with high-ionization 
narrow lines are classified as follows. Objects with
[Ne{\sc v}] detections with rest-frame 
equivalent width $>5$\AA$\;$are classed as type-2 with the basis column 
containing ``[Ne{\sc v}]'' and the measured rest-frame equivalent
width in Angstroms (for comparison 
the [Ne{\sc v}] 3426 emission line in the 
SDSS composite quasar spectrum of Vanden Berk 
et al. (2001) has an EW of $\approx 0.6$\AA). For low redshift objects
where H$\alpha$ was included in the spectra, we have used BPT diagrams
to classify the AGN.
Figure 4 shows the emission line ratios of 
our objects plotted on two diagnostic diagrams using the 
divisions between star-forming, Seyfert and LINER galaxies proposed
by Kewley et al.\ (2006). Most objects plot with the Seyferts, 
though two objects plot close to the AGN/starburst division line in both
plots, and are 
presumably composite starburst/AGN objects (SSTXFLS 171650.6+595752; 
SWIRE2 J022225.86-050015.1). We class these as type-2 as they do have evidence
of AGN activity, even if this is not dominating their optical 
emission line ratios. For some objects with 
neither [Ne{\sc v}] nor H$\alpha$/[N{\sc ii}] detected in their
spectra we have relied on the ionization parameter as an approximate
AGN indicator and
used the [O{\sc iii}]5007/[O{\sc ii}]3727 ratio, classing objects 
with [O{\sc iii}]5007/[O{\sc ii}]3727 $>1$ as questionable type-2s.
 
There are also a few high redshift objects having high-ionization 
UV permitted lines 
with widths $\sim 2000$kms$^{-1}$. Linewidths of this order are 
seen in the UV spectrum of NGC1068 (Grimes et al.\ 1999). 
Given this, and 
their high luminosities, we therefore anticipate that the true broad 
line widths
will be much higher, so we are probably seeing only the narrow-line region 
in our spectra. However, given their high redshifts ($z>1.5$), 
only a very small amount of extinction is required to hide the
broad-line region in the rest-frame UV. For the three highest redshift 
objects, at 2.78, 2.99 and 4.27, the rest-frame optical to near-infrared 
light is sampled in the IRAC bands, and this must almost certainly 
be dominated by reddened
quasar light for these objects to exceed our IRAC flux limits. It is therefore
unclear whether these objects should be classed as type-2s (showing no strong 
broad emission lines or quasar continuum in the rest-frame optical), or 
as class 1Rs. We thus assign $z>1.5$ objects with narrow UV lines  
class 2HZ (and group them with the type-2 objects for plotting and statistical
purposes), pending better information on their rest-frame optical 
emission line properties and SEDs from near infrared imaging and spectroscopy. 
In all of these objects, Ly$\alpha$ is typically faint compared to the 
high ionization UV lines, probably due to absorption by neutral hydrogen
(e.g.\ van Ojik et al.\  1994).

\subsection{Starbursts and LINERs}

We are left with seven objects with low-ionization lines, showing
no signs of obvious AGN activity. Such objects
have been classified as ``type-3'' AGN by Leipski et al.\ (2005). 
They are all relatively low redshift ($z<0.6$).
All show [O{\sc ii}]3727 in emission, and Balmer absorption features. One also
shows strong narrow H$\beta$ and H$\gamma$ emission. The two objects we
have been able to positively identify as starburst in the optical 
have been classified as such. The remainder are classified as
starbursts/LINERs. Note that we fail to find any unambiguous 
LINERs in our sample. (SSTXFLS 171454.4+584948 plots below the Seyferts in the
 N[{\sc ii}]/H$\alpha$ vs [O{\sc iii}]/H$\beta$ diagram, in the 
traditional LINER region [e.g.\ Kauffmann et al.\ 2003] but well within 
the starbursts in the [O{\sc i}]/H$\alpha$ vs [O{\sc iii}]/H$\beta$ diagram
which gives a better discrimination of LINERs according to Kewley et al.\
[2006], so we have classified it as a starburst.)
This may be due to LINERs lacking a 
substantial hot torus bright enough to dominate
the mid-infrared over stellar emission (e.g.\ Ogle et al.\ 2006), but 
better spectra of our starburst/LINER objects are needed to 
confirm this. 

For the most part our classifications are 
consistent with those of Papovich et al.\ (2006), though there are some objects 
which are differently classified. These are discussed in Appendix A.

\begin{table*}
{\scriptsize
\caption{Bright AGN in the XFLS region}
\begin{tabular}{lccccl}
Name &$S_{24.0}{\rm /mJy}^{1}$ & Redshift&Class&Basis\\ \hline
SDSS    J171033.1+584456 & 6.1 & 0.281&1R &Fe{\sc ii}/BL\\
SSTXFLS J171115.2+594906 & 9.4 &0.587&Sbt  &strong Balmer lines \\
SDSS    J171117.6+584123 & 5.8 &0.617&1   &BL\\
SDSS    J171124.2+593121 & 5.5 &1.491&1   &BL (LoBAL?)\\
SSTXFLS J171147.4+585839 & 4.8 &0.800&2  &[Ne{\sc v}]13\\
&&&&\\
SDSS    J171207.4+584754 &13.4 &0.269   &1 &Fe{\sc ii}\\
SSTXFLS J171233.4+583610 & 5.1 &1.663&1R &BL\\
SSTXFLS J171302.3+593610 &11.8 &0.668&2 &[Ne{\sc v}]33\\
SSTXFLS J171313.9+603146 &10.5 &0.124&2  &BPT\\
SSTXLFS J171324.2+585549 & 4.9 &0.609&2  &[Ne{\sc v}]12\\
&&&&\\
SSTXFLS J171325.1+590531 & 9.5 &0.126&2  &BPT\\
SSTXFLS J171331.5+585804 & 5.9 &0.435&Sbt/L  &[O{\sc iii}]/[O{\sc ii}]$<$1\\
SDSS    J171334.0+595028 & 5.4 &0.615&1   &BL\\
SSTXFLS J171335.1+584756 &23.7 &0.133&1R&Fe{\sc ii}\\
SSTXFLS J171345.5+600730 & 4.9 &2.78 &2HZ  &narrow UV lines\\
&&&&\\
SDSS    J171352.4+584201 &23.9 &0.521 &1 &BL/Fe{\sc ii}\\
SSTXFLS J171419.9+602724 & 5.6 &2.99&2HZ  &narrow UV lines\\ 
SSTXFLS J171430.7+584225 & 8.2 &0.561&1R  &BL$^{\dag}$\\
SSTSFLS J171454.4+584948$^*$ & 4.9 &0.253&Sbt  &BPT \\
SSTXLFS J171513.8+594638 & 5.0 &0.248&1R &BL (radio-loud)\\
&&&&\\
SSTXFLS J171530.7+600216 &11.6 &0.420&2 &[Ne{\sc v}]26\\
SSTXFLS J171650.6+595752 & 6.7 &0.182&2  &BPT (composite)\\
SSTXFLS J171708.6+591341 & 5.4 &0.646&2  &[Ne{\sc v}]37\\
SDSS    J171732.9+594747 & 4.6 &1.058   &1  &BL\\
SDSS    J171736.9+593011 & 6.4 & 0.599&1  &BL\\
&&&&\\
SDSS    J171747.4+593258 & 5.4 &0.248&1   &BL\\
SSTXFLS J171750.7+584745 & 5.1 &2.55?  &1R?  &BL?\\
SSTXFLS J171754.6+600913 & 8.9 &4.27&2HZ & narrow UV lines\\
SSTXFLS J171831.5+595317 & 8.3 &0.700&2  &[Ne{\sc v}]26\\
SSTXFLS J171839.6+593359 &10.9 &0.383&1 &BL (radio loud)\\
&&&&\\
SDSS    J171902.2+593715 &26.9 & 0.178  &1 &BL/Fe{\sc ii}\\
SSTXFLS J171913.5+584508 & 8.9 &0.318&2  &[Ne{\sc v}]12, BPT\\
SSTXFLS J172044.8+582923 & 5.1 &1.697&2HZ & narrow UV lines\\
SSTXFLS J172050.4+591511 & 9.4 &1.44    &1R &BL\\
SSTXFLS J172123.1+601214 &13.3 &0.325&2 &[Ne{\sc v}]33/BPT\\
&&&&\\
SSTXFLS J172219.5+594506 & 7.8 &0.271&2  &BPT\\
SSTXFLS J172228.1+601526 & 7.1 &0.741&2  &[Ne{\sc v}]15\\
SDSS    J172238.7+585107 & 6.7 & 1.617&1  &BL\\
SSTXFLS J172245.0+590328 & 4.7 &0.797&2  &[Ne{\sc v}]43\\
SSTXFLS J172248.9+583256 & 5.6 &0.84 & 1R & BL \\
&&&&\\
SSTXFLS J172328.4+592947 & 8.1 &1.34?&1R &BL\\
SSTXFLS J172542.3+595317 & 9.5 &0.437&Sbt/L & low ionization\\
SDSS    J172619.7+601600 & 6.6 & 0.925&1  &BL\\
\end{tabular}

\noindent
$^{*}$ Spectrum from Papovich et al.\ (2006).\newline
$^{\dag}$ classification based on Papovich et al.\ spectrum, see Appendix A.
$^{\ddag}$ Optical/near-IR redshift is uncertain, but a {\em Spitzer} mid-IR
spectrum shows silicate absorption at $z=0.84$ (Lacy et al.\ 2006, in prepraration).}
\end{table*}

\begin{table*}
{\scriptsize
\caption{Bright AGN in the SWIRE-XMM region}
\begin{tabular}{lcccl}
Name &$S_{24.0}{\rm /mJy}^{1}$ & Redshift&Class&Basis \\ \hline
SWIRE2 J021638.21-042250.8&14.6&0.304&2 &BPT\\ 
SWIRE2 J021640.72-044405.1&14.8&0.870&1 &Becker et al.\ (2001)\\ 
SWIRE2 J021657.77-032459.7&24.2&0.137&1 &BL\\ 
SWIRE2 J021729.06-041937.8&8.9&1.146&1R &BL\\ 
SWIRE2 J021749.00-052306.9&8.1&0.987?&Sbt/L?&low ionization?\\ 
&&&&\\
SWIRE2 J021759.91-052056.1&8.8&0.535&2  &[Ne{\sc v}]26\\ 
SWIRE2 J021808.22-045845.3&9.2&0.712&1  &Caccianiga et al.\ (2004)\\ 
SWIRE2 J021809.45-045945.9&7.8&1.094&2  &[Ne{\sc v}]90 \\ 
SWIRE2 J021830.57-045622.9&8.5&1.401&1  &Sharp et al.\ (2002)\\ 
SWIRE2 J021859.74-040237.2&16.0&0.203&2 &BPT\\ 
&&&&\\
SWIRE2 J021909.60-052512.9&25.6&0.099&2 &BPT \\ 
SWIRE2 J021938.70-032508.2&6.6&0.435&1  &Fe{\sc ii}\\ 
SWIRE2 J021939.08-051133.8&32.9&0.150&2 &[Ne{\sc v}]9\\ 
SWIRE2 J022005.93-031545.7&6.6&1.560&1  &BL\\ 
SWIRE2 J022012.21-034111.8&6.8&0.166&1  &BL\\ 
&&&&\\
SWIRE2 J022039.48-030820.3&10.6&0.451&1 &BL\\ 
SWIRE2 J022133.82-054842.8&6.9&0.502&1  &BL\\ 
SWIRE2 J022211.57-051308.1&6.7&0.841&2  &[Ne{\sc v}]30 \\ 
SWIRE2 J022225.86-050015.1&8.3&0.149&2  &BPT (composite)\\ 
SWIRE2 J022255.87-051351.7&8.3&0.846&1  &BL\\ 
&&&&\\
SWIRE2 J022301.97-052335.8&6.8&0.708?&Sbt/L?&low ionization?\\ 
SWIRE2 J022306.74-050529.1&16.0&0.330&2 &BPT \\ 
SWIRE2 J022310.42-055102.7&13.3&0.506&2 &[Ne{\sc v}]15\\ 
SWIRE2 J022348.96-025426.0&7.6&0.355&3  &low ionization\\ 
SWIRE2 J022356.49-025431.1&10.6&0.451&2 &[Ne{\sc v}]16\\ 
&&&&\\
SWIRE2 J022422.27-031054.7&7.8&1.226&1  &BL\\ 
SWIRE2 J022431.58-052818.8&9.4&2.068&1  &BL\\ 
SWIRE2 J022438.97-042706.3&6.6&0.252&2  &BPT \\ 
SWIRE2 J022455.77-031153.6&7.6&0.790&2? &[O{\sc iii}]/[O{\sc ii}]$>$1\\ 
SWIRE2 J022508.33-053917.7&9.6&0.293&2  &BPT\\ 
&&&&\\
SWIRE2 J022542.02-042441.1&10.9&0.580&2 &[Ne{\sc v}]23\\ 
SWIRE2 J022600.01-035954.5&9.0&1.095?&1R&BL\\ 
SWIRE2 J022612.67-040319.4&7.7&0.330&2  &BPT \\ 
SWIRE2 J022700.77-042020.6&27.1&0.053&1 &Fe{\sc ii}  \\
\end{tabular}
}
\end{table*}

\subsection{Optical to mid-infrared color}

In Figure 5 we show the observed $R$ - [8.0] color plotted against redshift
for the different spectral classifications. For the XFLS sample, the 
$R$ magnitudes are from the survey of Fadda et al.\ (2004), those for the
SWIRE sample are from the Cambridge Automatic Plate Measuring (APM)  
scans of the UK Schmidt Telescope plates (though in a few cases
of faint or blended sources the magnitudes have been estimated by eye from
digitized sky survey plates). Apart from a couple of 
low redshift AGN, which probably have their optical emission dominated
by their host galaxies, all the normal type-1 objects trace the expected 
color locus for type-1 quasars very well, particularly the ``IR luminous'' 
SED of Richards et al.\ (2006), suggesting that the mid-infrared
selection is not picking a very 
unusual population of type-1 quasars, but rather is finding
quasars with the properties  that we would
expect from a sample selected in the mid-infrared. After the 
host galaxy contribution to the $R$-band becomes negligible at 
$z\stackrel{>}{_{\sim}} 0.4$, the normal and obscured (type 1R, 2 and 3) 
quasars separate, with very few objects having moderately-red colors. This 
relatively clean separation shows that our optical spectral
classifications are indeed likely to be correct. 

The $z>1.5$ type-2 objects (type 2HZ in Table 3), which have intermediate UV 
linewidths and  whose precise optical classification is therefore 
unclear, are obviously redder than
the expected mid-infrared to optical color 
of normal quasars at these redshifts, suggesting that they are not simply 
quasars with unusually narrow lines, but do indeed have most of their 
nuclear emission extinguished by dust.

Our division between type-1 and
type-1R quasars also appears to be justified on the basis of this color
(at least for objects at $z\stackrel{>}{_{\sim}} 0.4$ 
where the host galaxy does not
affect the $R$-magnitude), with the type-1R quasars having rest-frame 
reddenings $E(B-V)\sim 0.3-0.7$.

\section{X-ray detections}

There is some overlap between objects in the SWIRE-XMM fields and 
archival {\em XMM-Newton} data. 21 of 34 sources in the SWIRE-XMM sample
fall on moderately deep (5-50ks) XMM exposures. These are sufficiently
deep to detect, or to set useful limits on, the X-ray flux from 
our sources.
These data appear to have been taken for several different surveys, 
including the {\em XMM} large-scale structure survey (e.g.\
Chiapetti et al.\ 2005) and the
the Subaru {\em XMM-Newton} Deep Survey (SXDS; Watson et al.\ 2004). 
We have obtained approximate fluxes for these objects 
in order to investigate the optical-to-X-ray luminosity ratios of our 
sources (Table 5). This allows us to place our mid-infrared selected obscured 
AGN into the context of X-ray samples of similar objects.
Counts in the 0.2-12keV range 
from the PN detector of the European Photon Imaging Camera (EPIC), as 
listed in the 1XMM serendipitous 
catalog (e.g.\ Michel et al.\ 2004), or the pre-release of the 2XMM 
catalogue (The Second XMM-Newton Serendipitous Source Pre-release Catalogue,  
XMM-Newton Survey Science Centre) have been used where available,
otherwise counts have been taken from the source lists
available from the XMM archive (the pipeline products [PPS] lists).
Limits are estimated from the fluxes
of the faintest believable detections on each image, and correspond
to $\approx 3\sigma$.
Fluxes from the 1XMM or 2XMM catalogs are from the PN countrate in 1XMM, and
from the average over all detectors in 2XMM. Where only counts or a limit 
are available fluxes are derived from the count rate
using the PIMMS website (http://heasarc.gsfc.nasa.gov/Tools/w3pimms.html),
assuming an input spectrum with photon index of two and a column of 
$10^{21} {\rm cm^{-2}}$.

A few points are clear from even this preliminary study. All the 
type-1 quasars are detected in the X-ray (albeit with about an order
of magnitude of scatter in their X-ray to infrared ratios), 
but only about half of the 
optically-obscured quasars (Figure 6). The obscured quasars do, 
however, show a wide range in X-ray
to infrared flux ratios, ranging from little or no absorption to 
X-ray fluxes absorbed by
a factor of ten or more (averaged over the 0.2-12keV band) 
relative to the type-1s. 
Interestingly, one of the two starburst/LINER objects in these X-ray fields is 
detected, though the other is not.

\begin{figure}

\plotone{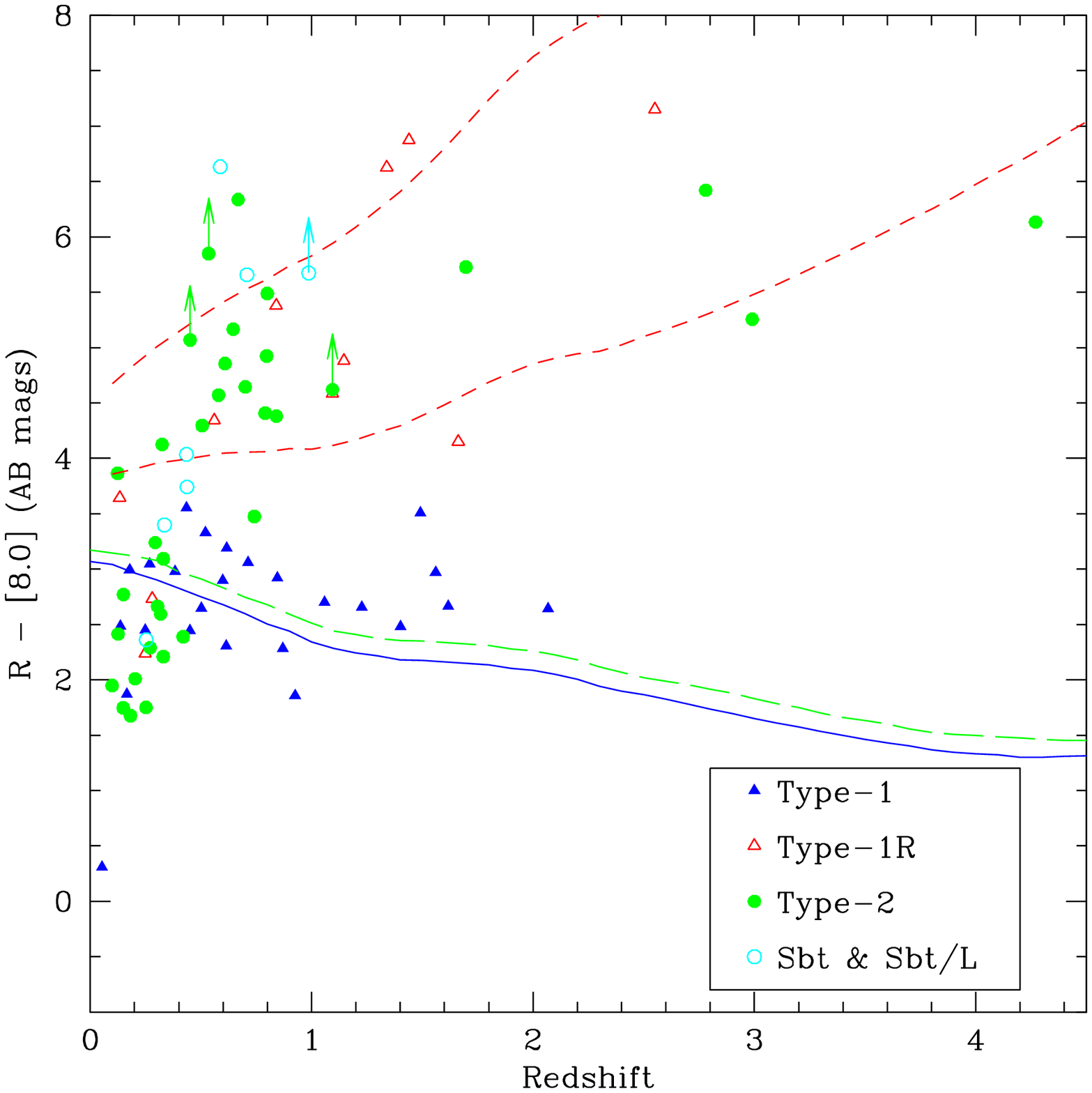}

\caption{Optical to infrared color for our quasar samples. 
Type-3 objects which lack
redshifts have been assigned a redshift of 1.50. The color loci of  
two of the quasar SEDs of Richards et al. (2006) are plotted as lines. The
blue solid line is the mean quasar SED and the green long-dashed line the 
``IR luminous'' SED. The red dashed lines indicate the colors of the mean 
quasar
SED reddened by the Small Magellenic Cloud extinction law of Pei (1992). The
lower line corresponds to a  rest-frame reddening of $E(B-V)=0.32$ and
the upper to $0.65$. The dispersion in optical to 
mid-IR color of the quasars of Richards et al.\ (2006) is $\approx \pm 0.3$
mag.}

\end{figure}

\begin{figure}

\plotone{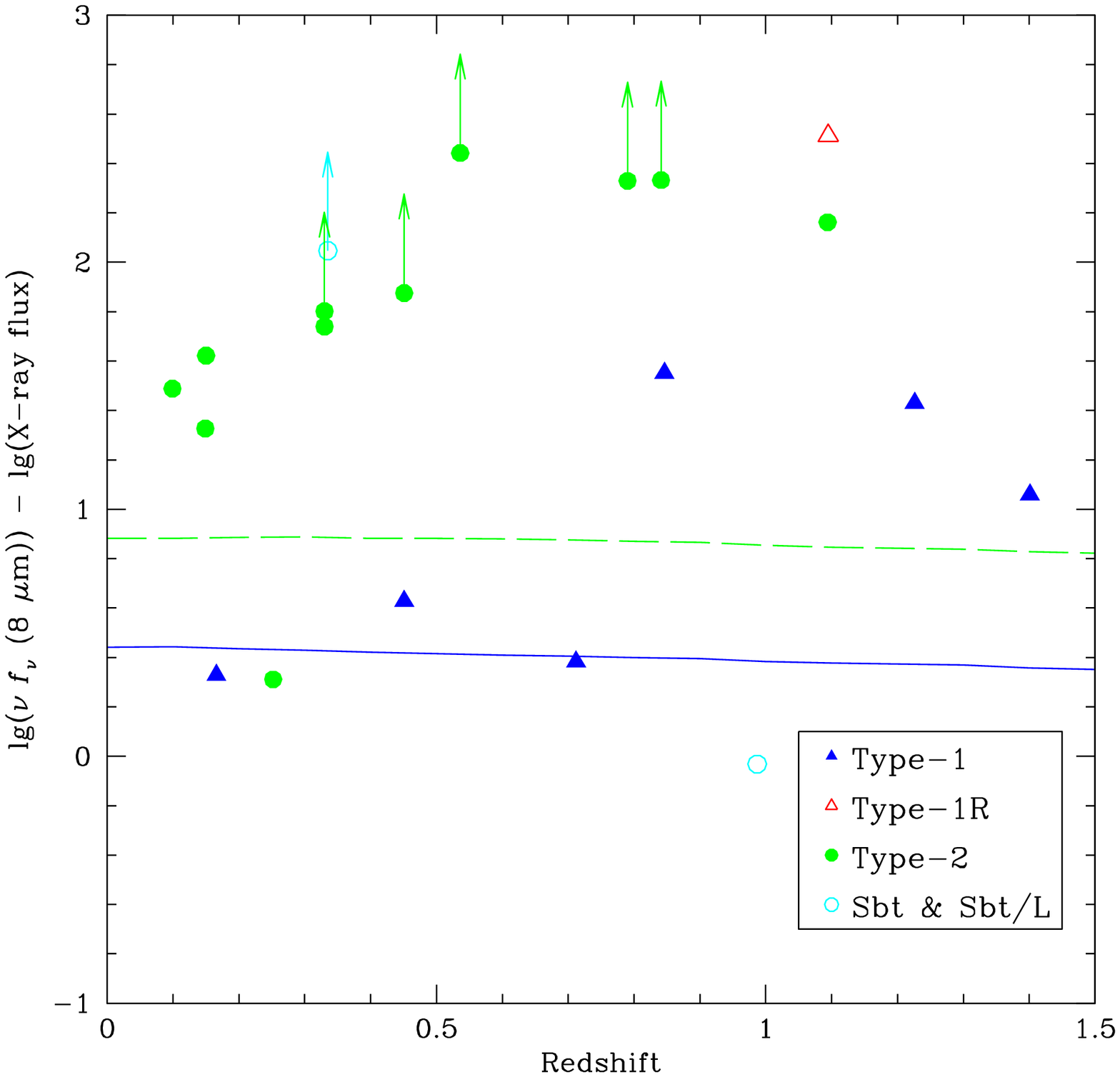}

\caption{Observed
X-ray (0.2-12 keV) flux  
versus observed 8.0$\mu$m infrared flux ratios for objects in archival XMM fields. 
The blue solid line is the mean
quasar SED of Richards et al.\ extrapolated from the soft X-ray assuming 
a photon index of 1.8, the green dashed line the same for the ``IR luminous'' 
SED.}

\end{figure}

\begin{figure}

\plotone{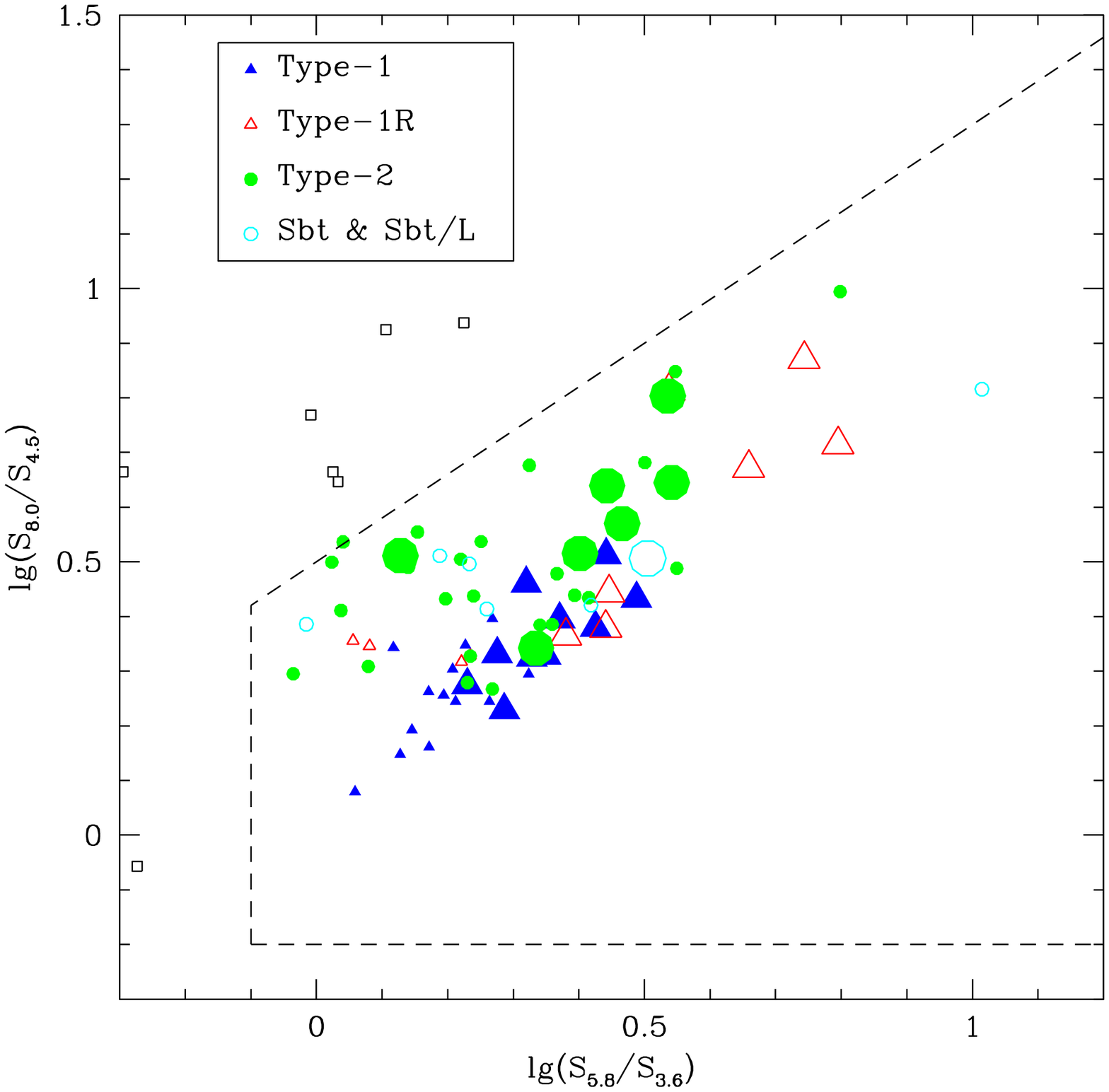}

\caption{Enlarged view of the ``AGN wedge'' of Figure 1. Large symbols 
represent $z\geq 0.8$ objects, small symbols those at $z<0.8$. Open squares
represent the objects from the sample of Table 6 which have AGN-like optical 
emission line ratios but non-AGN IRAC colors.}

\end{figure}

\begin{figure}

\plotone{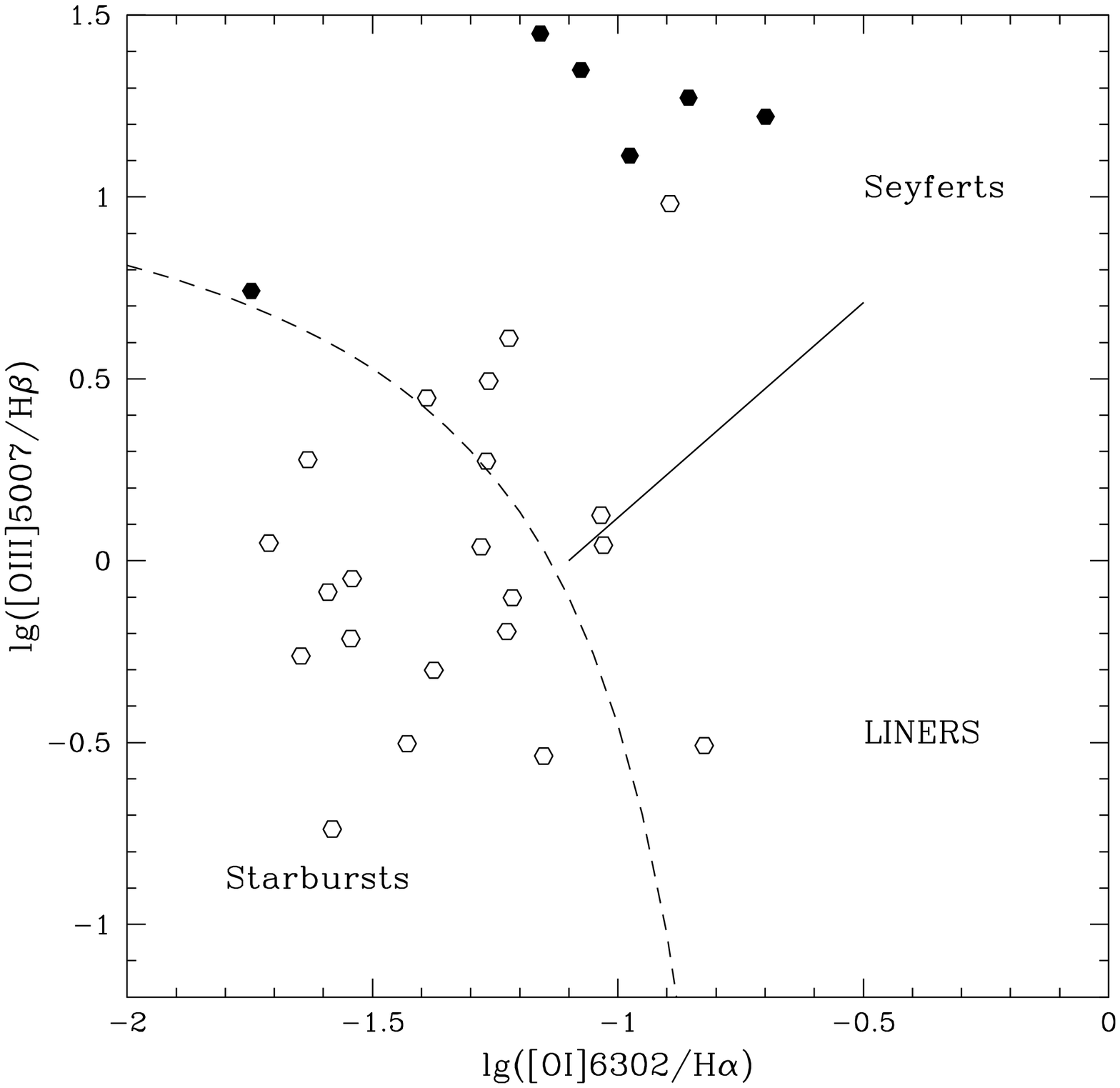}

\caption{[O{\sc i}]/H$\alpha$ vs [O{\sc iii}]/H$\beta$ 
BPT diagram for the objects of 
Table 6 (the 50 brightest 24$\mu$m sources, regardless of 
IRAC colors) with [O{\sc iii}] detections. Objects falling outside of
the IRAC AGN color selection are shown as open symbols, those 
within it as closed.}

\end{figure}

\section{Discussion}

\subsection{Selection effects and redshift biasses}

Every technique for selecting AGN is affected by selection biases, 
and this one is no exception.  The color selection 
means that objects whose observed mid-infrared colors are not dominated by thermal emission from 
the AGN will be missing from the sample. This can happen in a variety of ways. First, the AGN may 
simply be of such low luminosity in the infrared that the starlight from the host galaxy dominates
the mid-infrared emission, either because the AGN is intrinsically weak, or because the covering 
factor of the hot dust is low, perhaps because of an unusual torus geometry. 
This bias gets worse with increasing redshift as the stellar bump in the rest-frame 
near-infrared is redshifted into the mid-infrared bands, and the $k$-correction 
on the thermal emission of the dust becomes large. Second, some AGN will simply be so highly-obscured 
that the AGN emission is not seen at observed wavelengths $<8\mu $m. Third, some AGN 
may also be so highly obscured that they drop out of the IRAC catalogs at short wavelengths. 

The positions of the different classes of objects within the ``AGN wedge'' on the IRAC color-color plot of Figure 1, shown in detail in Figure 7,
can give us some idea of the particular selection effects for each class of object. The type-1 quasars
all lie in a fairly tight group in the center of the wedge, so few are likely to be missing 
(see also Paper 1). Reddened type-1s show more dispersion, including two with 
very red IRAC colors.
The type-2s are also more dispersed, with several (mostly low redshift
objects) trending towards the diagonal divide in the 
color-color plot, presumably due to PAH emission, and several with very red colors. 
The starbursts and starburst/LINERs are surprisingly close to the center of the
wedge, though one does have extremely red IRAC colors. 
From Figure 7 we can deduce that some (probably mostly lower-$z$)
type-2 objects are probably missing due to plotting with the PAH-dominated galaxies, and that all classes
of obscured quasars are probably missing their reddest representatives. There is a weak trend 
whereby the higher redshift objects tend to be located towards redder colors in the wedge, though
at the highest redshifts, where the IRAC colors are sampling the rest-frame optical and near-infrared,
the colors become bluer again.

As these selection effects have yet to be modeled in depth, we will defer further discussion to a later paper, where we will attempt to model the emission from 
our objects in detail and apportion luminosity between AGN and starburst 
components (Petric et al.\ 2006).
However, we point out
here that our fraction of obscured quasars is necessarily a lower limit, and this limit becomes
increasingly poor with increasing  redshift.

\subsection{Objects with low ionization optical spectra}

Are there any high luminosity AGN in which both the broad and narrow-line 
regions are obscured, or otherwise missing? This study has found several 
objects which qualify as ``type-3'' quasars (following 
Leipski et al.\ [2005]), namely galaxies with emission line 
properties more typical of LINER or starburst galaxies than narrow-line AGN. 
Leipski et al.\ find 57\% of their ISO/2MASS selected AGN have these optical 
spectra. Mart\'\i nez-Sansigre et al.\ (2005, 2006) also deduce 
the presence of a significant population of  obscured narrow-line AGN  from 
their mid-infrared/radio selected objects which show no emission lines in 
the rest-frame UV. 
 
Our study has a lower fraction. Including both objects with starburst or 
ambiguous starburst/LINER spectra we find a ``type-3'' fraction of 
$\approx 9$\%, which is of course an upper limit as, apart from 
one case which is detected in the X-ray at the level 
expected for an AGN (SWIRE2 J021749.00-052306.9), we cannot prove the 
presence of AGN in these objects. Thus, despite the inferred presence of 
hot dust, infrared spectroscopy or X-ray detection (at a level greater than
expected for a starburst) is required to prove the 
AGN nature of these candidate type-3 objects. If they are reddened AGN, the 
reddening may be in the host (e.g.\ Mart\'\i nez-Sansigre et al.\ 2005; Rigby 
et al.\ 2006), though we cannot eliminate the possibility that nuclear dust 
is obscuring the narrow-line region, or is blocking the line of 
sight to the narrow line region from the nucleus.

X-ray surveys too find a population of AGN which lack opical evidence for AGN 
activity, the X-ray bright, optically-normal galaxies (XBONGs)
(e.g.\ Comastri et al.\ 2002). A number of processes have been suggested for 
the lack of optical AGN activity in these objects, including host galaxy 
obscuration (Rigby et al.\ 2006), radiatively inefficient accretion 
(Yuan \& Narayan 2004) or dilution of weak emission lines by host galaxy light
(Moran, Filippenko \& Chornock 2002; Eckhart et al.\ 2006). The discovery of 
[N{\sc v}]3426 emission in a stacked spectrum of XBONGS made by Eckart et 
al.\ (2006) favors dilution and/or 
extinction over a fundamental difference in accretion mode. 
The same may well be true if our type-3 quasars really are AGN, particularly 
any that are composite starburst/AGN systems with strong star formation 
activity in the host galaxies. In particular, objects which are classified on 
the basis of line ratios in the BPT diagram may have their AGN line emission 
overwhelmed by the starburst component, which is likely to dominate
the overall energetics of the galaxy. 

Dilution of AGN signatures by host galaxy emission may
help to explain why our type-3 fraction is lower
than those of Leipski et al., whose objects are mostly of lower redshift 
and luminosity, where host galaxy contamination is more of an issue. Our 
type-3 fraction may be smaller than that of
Mart\'\i inez-Sansigre et al.\ (2005,2006) due to our order of magnitude 
brighter flux limit at 24$\mu$m allowing us to 
find more objects with intrinsically faint or lightly-reddened 
($E(B-V) \sim 0.1-1$) narrow lines. Certainly it is 
apparent from our spectroscopic campaign that even objects with 
24$\mu$m fluxes of several mJy can
require 1.5hrs of integration on a 5m telescope to see emission lines. 

\begin{table*}
\caption{XMM-Newton detections of AGN in the SWIRE field}
{\scriptsize
\begin{tabular}{lcccccccc}
Object name &$z$&Opt& Obs ID & Exposure & Camera & counts& source & $S_{0.2-12{\rm kev}}$\\
            &&class& name      & time     &        &    &        & (10$^{-14}$\\
                          &      &  &          &     &     &      &   & ergs$^{-1}$cm$^{-2}$) \\ \hline
SWIRE2 J021749.00-052306.9&0.987?&Sbt/L??&0112370701&47648&PN   &20050 &2XMM& 178 \\
SWIRE2 J021759.91-052056.1&0.535 &2&0112370801&47330&PN   &$\stackrel{<}{_{\sim}} 50$&PPS&$\stackrel{<}{_{\sim}}$ 0.4\\
SWIRE2 J021808.22-045845.3&0.712&1&0112370101&57631&PN   &14500 &1XMM&104\\
SWIRE2 J021809.45-045945.9&1.094&2&0112370101&57631&PN   &100&PPS &0.75\\
SWIRE2 J021830.57-045622.9&1.401&1&0112370101&57631&PN   &3100 &1XMM&22\\
                          &     &            &     &     &       &   & \\
SWIRE2 J021909.60-052512.9&0.099&2&0112370801&47330&PN   &300&2XMM&2.6\\
SWIRE2 J021939.08-051133.8&0.150&2&0112370801&47330&PN   &601&1XMM&5.3\\
SWIRE2 J022012.21-034111.8&0.166&1&0037982601&11592&PN&1740&2XMM&63\\
SWIRE2 J022039.48-030820.3&0.451&1&0037982701&14892&PN&1100&2XMM&31\\
SWIRE2 J022211.57-051308.1&0.841&2&0111110501&20056&PN&$\stackrel{<}{_{\sim}} 20$&PPS&$\stackrel{<}{_{\sim}} 0.4$\\
                          &     &            &     &     &       &   & \\
SWIRE2 J022225.86-050015.1&0.149&2&0111110501&20056&MOS2&30&PPS&2.0\\
SWIRE2 J022255.87-051351.7&0.846&1&0111110501&20056&PN&20&1XMM&22\\
SWIRE2 J022306.74-050529.1&0.330&2&0111110501&20056&PN&46&1XMM&0.95\\
SWIRE2 J022348.96-025426.0&0.355&Sbt/L&0037981601&13319&PN&$\stackrel{<}{_{\sim}} 26$&PPS&$\stackrel{<}{_{\sim}} 0.8$\\
SWIRE2 J022356.49-025431.1&0.451&2&0037981601&13319&PN&$\stackrel{<}{_{\sim}} 26$&PPS&$\stackrel{<}{_{\sim}} 0.8$\\
                          &     &            &     &     &       &   & \\
SWIRE2 J022422.27-031054.7&1.226&1&0037981201 &8420 &PN&170&2XMM&8.4\\
SWIRE2 J022438.97-042706.3&0.252&2&0112680501&10040 &PN&1600&2XMM&30\\
SWIRE2 J022455.77-031153.6&0.790&2?&0037981301&5702 &PN&$\stackrel{<}{_{\sim}} 10$&PPS&$\stackrel{<}{_{\sim}} 0.8$\\
SWIRE2 J022542.02-042441.1&0.580&2&0112681001&41946&PN&$\stackrel{<}{_{\sim}} 442$&PPS&$\stackrel{<}{_{\sim}} 0.4$\\
SWIRE2 J022600.01-035954.5&1.095?&1R&0112680201&13675 &PN&33&2XMM&1.0\\
                          &     &            &     &     &       &   & \\
SWIRE2 J022612.67-040319.4&0.330&2&0112680201&13675&PN&$\stackrel{<}{_{\sim}} 27$&PPS&$\stackrel{<}{_{\sim}} 0.8$\\
\end{tabular}
}
\end{table*}

\begin{table*}
\caption{The 50 brightest extragalactic 24$\mu$m sources in the XFLS with 4-band IRAC coverage}
{\scriptsize
\begin{tabular}{lccccccc}
Name             &  $S_{24}$ (mJy) & $R$ & Spectrum$^{\ddag}$&Redshift & Class & Basis & IRAC-selected AGN?\\ \hline
SSTXFLS J171115.2+594907 &  9.41 & 20.97 & L06& 0.587 & Sbt & strong Balmer lines & yes\\
SSTXFLS J171150.2+590041 & 10.88 & 17.97 & P06&0.060 & 2 & BPT & no\\
SDSS J171207.4+584754    & 13.34 & 17.76 & L06&0.269 & 1 & Fe{\sc ii} & yes\\
SSTXFLS J171232.3+592126 &  8.47 & 17.56 & P06&0.21  & Sbt/L & low ionization&no\\
SSTXFLS J171302.3+593611 & 11.83 & 21.50 & L06&0.668 & 2 &[Ne{\sc v}]33 &yes\\
                         &       &       &       &    &     & &\\
SSTXFLS J171313.9+603146 & 10.49 & 18.31 & L06&0.105 & 2  &BPT &yes\\
SSTXFLS J171325.1+590531 &  9.43 & 18.33 & L06&0.126 & 2  &BPT & yes\\
SSTXFLS J171335.1+584756 & 23.66 & 17.56 & L06&0.133 & 1R & Fe{\sc ii} & yes\\
SDSS J171352.4+584201 & 23.90 & 17.53 & SDSS&0.521& 1 & BL/Fe{\sc ii} & yes \\
SSTXFLS J171414.8+585221 &  8.97 & 18.70 & P06&0.167 &Sbt & BPT& no\\
                         &         &     &       &   &      &    & \\
SSTXFLS J171430.7+584225 &  8.24 & 19.72 & L06&0.561 & 1R & BL & yes\\ 
SSTXFLS J171437.4+595647 & 11.60 & 18.75 & P06&0.196 & 2 & BPT (composite)& no\\
SSTXFLS J171446.4+593359 &  7.37 & 17.67 & P06&0.129 & Sbt/L & low ionization&no\\
SSTXLFS J171530.7+600216 & 11.55 & 18.82 & L06&0.420 & 2 & [Ne{\sc v}]26 & yes \\
SDSS J171540.1+591647 &  6.88 & 17.43 & SDSS&0.116 & Sbt/L & low ionization&no\\ 
                         &         &     &       &  &       &    & \\
SSTXFLS J171542.0+591657 & 26.33 & 17.50 &P06& 0.116 &Sbt & BPT & no\\
SSTXFLS J171544.0+600835 &  6.86 & 17.07 &L06$^{*}$& 0.157 & 2 & BPT & no\\  
SDSS J171607.2+591456    & 34.57 & 16.27 &SDSS& 0.054 & Sbt & BPT & no\\
SDSS J171614.4+595423    &  8.41 & 17.39 &SDSS& 0.153 & Sbt & BPT & no\\
SDSS J171630.1+601423    &  8.35 & 17.82 &SDSS& 0.107 & Sbt & BPT & no\\
                         &         &     &       & &        &    & \\
SSTXFLS J171634.0+601443 & 10.76 & 17.04 &P06& 0.108 & Sbt & BPT & no\\
SDSS J171635.9+601436    &  7.66 & 17.07 &SDSS& 0.106 &Sbt & BPT & no\\
SDSS J171641.0+591857    & 20.88 & 16.41 &SDSS& 0.056 &Sbt & BPT & no\\
SSTXFLS J171650.5+595751 &  6.69 & 18.81 &L06& 0.182 & 2 & BPT (composite)& yes\\
SSTXFLS J171711.1+602710 &  9.43 & 17.88 &L06$^{*}$& 0.110& Sbt/L & low ionization& no\\
                         &         &     &       &&         &    & \\
SSTXFLS J171744.1+583848 & 13.92 & 17.73 &P06& 0.066 &L& BPT &no\\
SSTXFLS J171754.6+600913 &  8.94 & 21.80 &L06& 4.27  &2HZ & narrow UV lines & yes\\
SSTXFLS J171831.7+595317 &  8.27 & 20.56 &L06& 0.700 & 2  & [Ne{\sc v}]26 & yes\\
SSTXFLS J171839.7+593359 & 13.22 & 17.99 &L06& 0.383 & 1 & BL & yes\\
SSTXFLS J171852.7+591432 & 13.92 & 19.23 &P06& 0.322 & 2/L& BPT &no\\
                         &         &     &       &&         &    & \\
SSTXFLS J171902.2+593715 & 26.91 & 17.10 &L06& 0.178 & 1 & BL/Fe{\sc ii} & yes\\
SSTXFLS J171913.5+584509 &  8.90 & 19.00 &L06& 0.318 & 2 & [Ne{\sc v}]12, BPT &yes\\
SSTXFLS J171916.6+593449 & 17.40 & 17.91 &P06& 0.166 & 2/L& BPT &no\\
SSTXFLS J171933.3+592742 &  7.53 & 17.71 &P06& 0.139 &2& BPT (composite) & no\\
SDSS J171944.8+595705$^{\dag}$ &  6.73 & 16.66 & SDSS&0.069 &Sbt & BPT & no\\
                         &         &     &      & &         &    & \\
SDSS J171944.9+595707$^{\dag}$& 7.44    & 16.66&SDSS& 0.069 & Sbt & BPT &no\\
SSTXFLS J172028.9+584749 &  9.57 & 16.74 & L06$^{*}$& 0.127 & 2 & BPT & no\\
SDSS J172043.2+584026    &  9.48 & 17.08 &SDSS& 0.125&Sbt & BPT & no\\
SSTXFLS J172050.5+591512 &  9.41 & 21.62 &L06&1.44 & 1R & BL & yes\\
SSTXFLS J172123.1+601214 &  13.34& 18.78 &L06&0.325& 2& [Ne{\sc v}]33, BPT &yes\\
                         &         &     &     &  &         &    & \\
SSTXFLS J172219.5+594507 &   7.76  & 18.71 &L06& 0.272 & 2 & BPT & yes\\
SSTXFLS J172220.2+590949 &   8.78  & 17.58 &P06& 0.179 & Sbt& BPT & no\\
SSTXFLS J172228.0+601525 &  7.12   & 19.85 &L06& 0.741 &2 & [Ne{\sc v}]15&yes\\
SDSS J172238.7+585107 &   6.73  & 18.17 &SDSS& 1.617 & 1 & BL & yes\\ 
SSTXFLS J172328.4+592947 &  8.10&  22.15&L06& 1.34? & 1R & BL & yes\\
                         &         &     &    &   &         &    & \\
SSTXFLS J172400.6+590228 & 13.34   & 17.99 &P06&  0.178& Sbt & BPT & no\\
SDSS J172402.1+600601  & 7.76 & 17.12 & SDSS &0.156& Sbt/L & low ionization&no\\
SSTXFLS J172508.8+591016 &  9.40 & 17.78 &P06& 0.164& Sbt & BPT & no\\
SSTXFLS J172542.3+595317 &  9.45 & 20.03 &L06& 0.437 & Sbt/L & low ionization & yes\\
SSTXFLS J172619.7+601559 &   6.57&  17.13&L06& 0.925 & 1 & BL & yes\\
\end{tabular}

\noindent
$^{\ddag}$Provenance of spectra: L06 - this paper, P06 - Papovich et al.\
(2006), SDSS - Sloan Digital Sky Survey data release 4.

\noindent
$^{*}$ Additional spectra obtained using COSMIC on the Palomar 200-inch, 2005, June 27-28

\noindent
$^{\dag}$ Parts of a single galaxy
}
\end{table*}

\subsection{Reliability and effectiveness of mid-infrared selection of AGN}

The IRAC color technique is clearly effective at selecting AGN, at 
least for objects with fluxes of a few mJy at 24$\mu$m.
We have been able to unambiguously classify most of our objects as AGN based
on their optical spectra. Contaminants are certainly possible, however. 
At moderate redshift, starbursts will be able to move into the AGN ``wedge''
if they have large amounts of warm dust, and at high redshifts, the 1.6$\mu$m
bump of stellar SEDs can lead to IRAC colors which mimic those of AGN. 
We can state that for this sample of 77 bright 24$\mu$m sources, only seven
lack obvious 
optical signatures of AGN activity, and of these, one is detected in 
an XMM field, which implies a contamination rate of $<$ 6/77 (8\% ). 

Of course, the selection of bright 24$\mu$m sources itself produces a large
fraction of AGN, as demonstrated by Brand et al.\ (2006). In particular, 
when combined with optical photometric criteria, a sample containing
a large fraction of AGN can be made by simply rejecting the optically-bright
objects. To investigate how effective such a cut is, and to take a first
look at the completeness of our AGN sample, 
we have examined the brightest 50 sources in the XFLS 24$\mu$m list (Table 6).
This sample contains 29 objects which are classed as AGN according to the 
criteria of Section 3. As the starbursts are also concentrated to 
lower redshifts, this qualitatively confirms the result of Brand et al.\ 
(2006), namely that the mid-infrared luminosity density is dominated by
AGN at mJy flux levels. Objects with faint optical identifications
are certainly more likely to be AGN - 
no object with $R>21$ is classified as a starburst or LINER, but
apart from that division there is a wide range in magnitudes 
($R\approx 17-20$) over which AGN and starburst/LINERs overlap, particularly
if no optical morphology criterion is used and thus quasars are included. 

This sample of the 50 brightest 24$\mu$m sources 
also allows us to take a first look at how complete the AGN selection 
using IRAC colors is. Interestingly, we find that seven objects outside the
IRAC color-based AGN selection show AGN-like emission line ratios in a 
BPT diagram, though these are nearly all close to the AGN-starburst
transition region (Figure 8). Their positions in the IRAC color-color 
plot are shown in Figure 7. In six of these cases IRAC 8.0$\mu$m
band flux is very high, pushing the objects into the starforming galaxy
region of the color-color plot of Figure 1. The remaining object 
(SSTXFLS J171150.2+590041) shows pure Seyfert emission line ratios, but 
 has only a weak mid-infrared excess and 
is in the part of Figure 1 dominated by quiescent galaxies. 
This suggests that in some cases the signature of AGN activity in the 
mid-infrared can be overwhelmed by strong star formation activity, 
at least at $z\stackrel{<}{_{\sim}}0.3$ where the PAH features can still 
contribute to the IRAC channel 4 flux. It may also indicate that much 
star formation activity (particularly any close to the nucleus) 
may be more obscured by dust than the narrow-line region of an AGN in the
same galaxy. An AGN selection with the diagonal
cut of the wedge raised by $\approx +0.5$ in ${\rm lg}(S_{8.0}/S_{5.8})$ would 
therefore increase the completeness of the AGN selection at low redshifts, 
though at the risk of reducing reliability.

\begin{figure}

\plotone{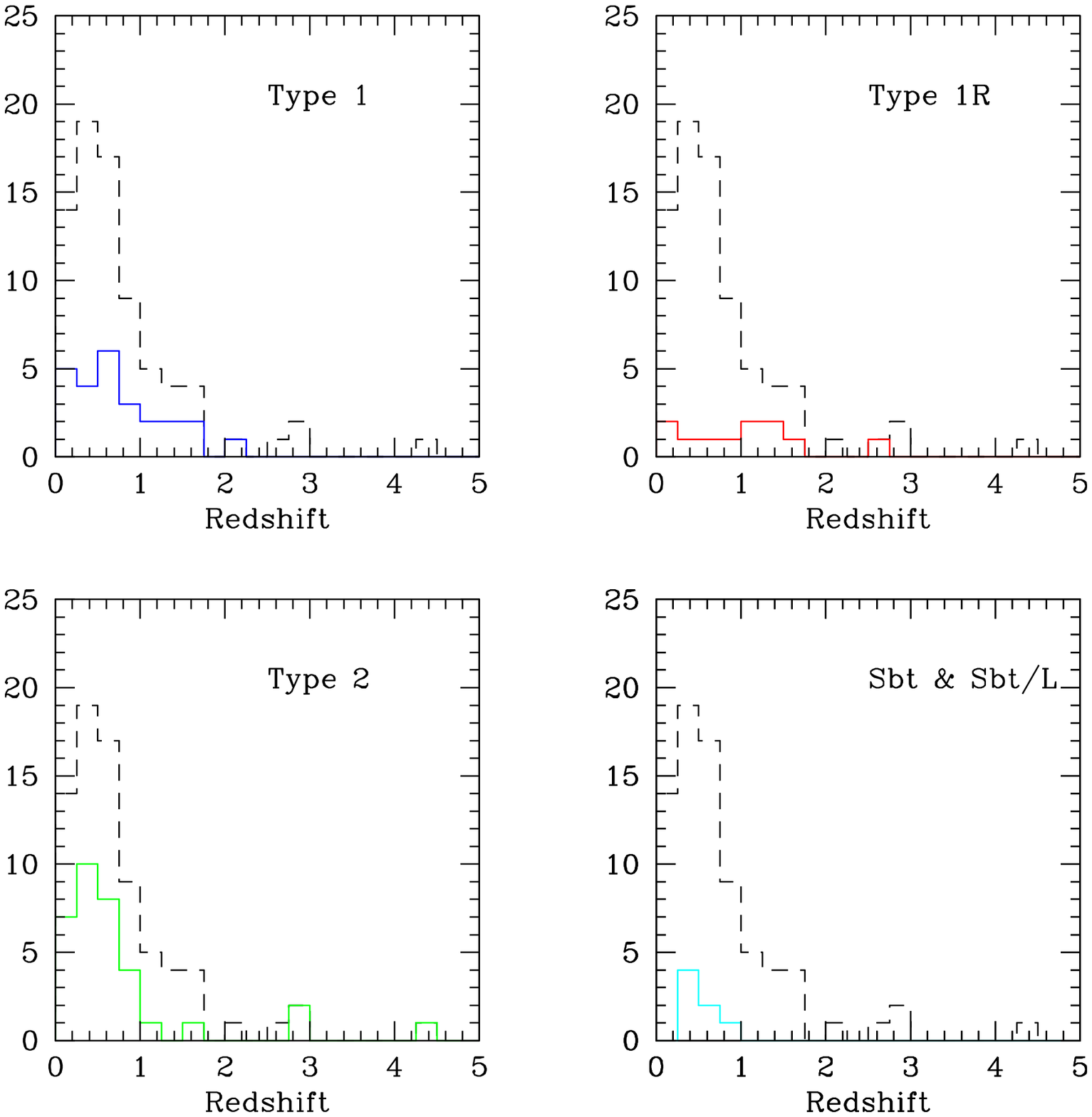}

\caption{The redshift distribution for our different classes of object.
The total distribution for the whole sample is shown as the dotted line on 
all the plots, and the solid line the distribution by type.}

\end{figure}

\begin{figure}

\plotone{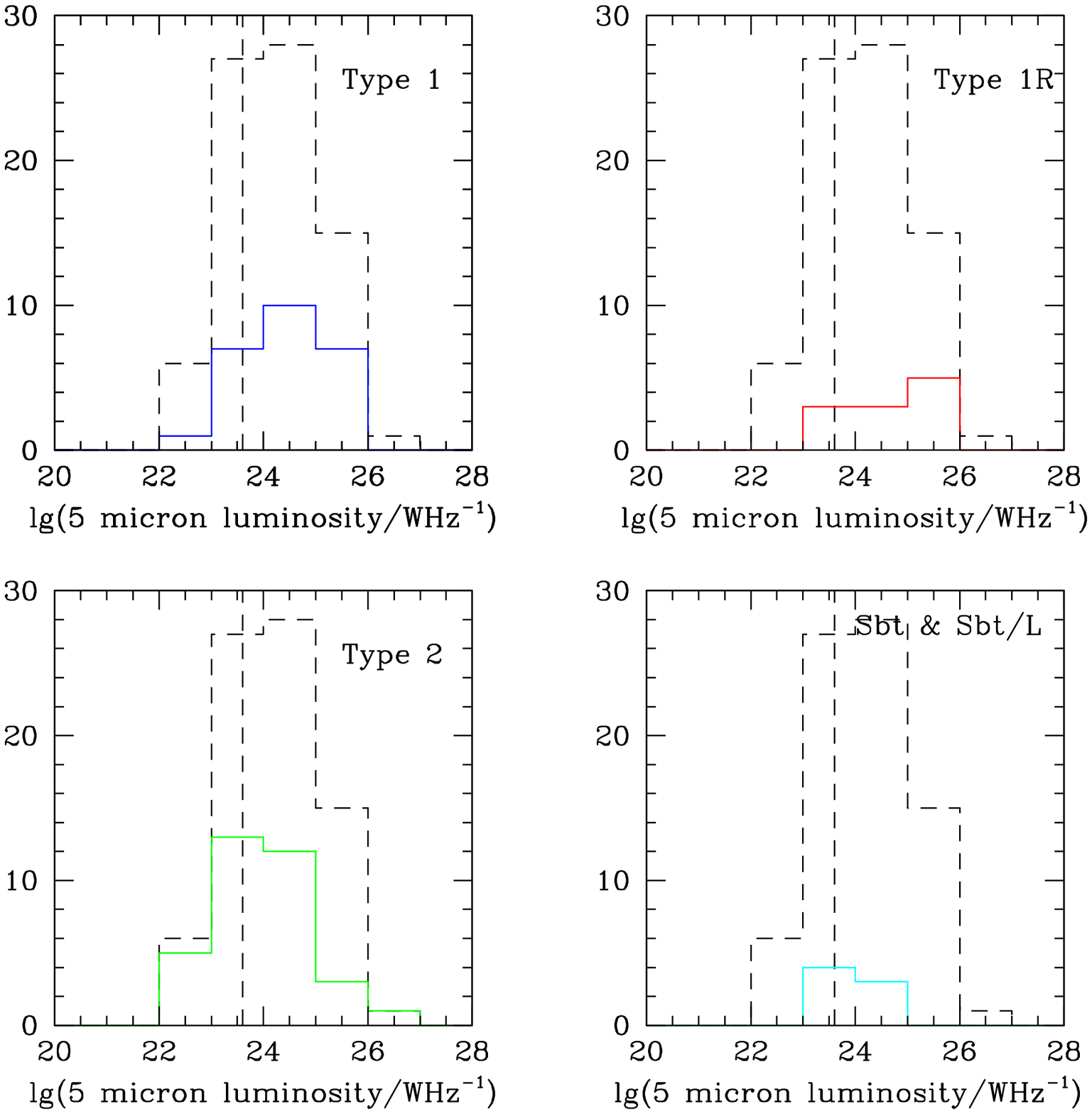}

\caption{As Figure 6, but for rest-frame luminosity at 5$\mu$m. The vertical
dashed line is at the approximate division between Seyfert and quasar 
luminosities.}

\end{figure}

\section{The ratio of obscured to unobscured AGN}

\subsection{Redshift and luminosity distributions}

Figure 9 shows the redshift histograms for our various classes of objects, 
compared to the redshift distribution of the sample as a whole. 

Figure 10 shows the corresponding luminosity distributions of our objects.
We calculate luminosities at a rest-wavelength of 5$\mu$m. This region
of the continuum is free from features, and can be calculated by 
interpolation between measurements 
over our full range of redshifts. The most heavily 
reddened AGN will, however, still be significantly extincted at this 
wavelength, which is a possible bias in Figure 10. 

As previously discussed, there are strong selection effects favoring the 
selection of low-$z$, high luminosity AGN in our sample. Therefore we have
made no attempt to investigate in detail the ratio of obscured to unobscured
AGN as a function of redshift or luminosity to compare with, for example, 
the receding torus model of Lawrence (1991).

Overall, the breakdown into optical spectral types is as follows:
33\% (25/77) normal type-1, 14\% (11/77) reddened type-1, 44\% (34/77) 
type-2 (including 2HZ), and 9\% (7/11) starbursts and ambiguous 
starbursts/LINERs. Defining an obscured quasar as anything with 
$E(B-V) \stackrel{>}{_{\sim}}0.3$ in the rest frame and 
an AGN nature confirmed through optical emission lines or X-rays 
(the type-1Rs,
type-2s, and the single X-ray detected starburst/LINER) we obtain a ratio of 
obscured to unobscured objects of 1.8:1. If we assume that the other six
starburst and starburst/LINER objects also contain obscured AGN this 
increases to 2.1:1. 

The fraction
of type-1Rs is higher than that found by Brown et al.\
(2006) in their 24$\mu$m selected sample. Brown et al find 20\%
of their broad-line objects are classed as reddened type-1s, compared to a 
type-1R/(type-1$+$type-1R) fraction of 30\% (14/46) in our sample. Our deeper
spectroscopy limit ($R\sim 23$ compared to $R<21.7$) is probably responsible
for this difference. Two of our type-1R quasars have $R<21.7$, and a third
has $R=21.6$, excluding these would reduce our type-1R fraction 
to 25\%. 

Before we can relate these ratios to an estimate of how much supermassive
black hole growth is obscured, we need to bear in mind the intrinsic 
variations in quasar SEDs. Selection in the mid-infrared will inevitably
mean our sample is biased in favor of objects whose intrinsic SEDs 
(before reddening of the quasar nucleus) are particularly bright in the 
mid-infrared, e.g.\ because of an unusually large dust covering 
factor. For type-1 and type-1R objects we have good evidence that this 
is not a large effect. Brown et al.\ (2006) find that the space density of 
their infrared-selected quasars is very similar to that of optically-selected
samples, 
consistent with the small systematic variations seen in the quasar SEDs
of Richards et al.\ (2006) and Hatziminaoglou et al.\ (2005) between 
optically-bright and infrared-bright quasars. Although our type-1 quasars
at $z>1$ are redder in $R$-[8.0] color than the Richards et al.\ template 
color in Figure 5, most are close to the template, 
and several at $z<1$ are bluer. For the more heavily obscured
objects there are two arguments that can be brought to bear. First, even 
in type-1 quasars about 40\% of the UV emission is reprocessed by dust (Richards et al.\ 2006). 
Even if the obscured quasars have dust covering factors near unity, their
dust emission would only increase by a factor $\sim 2$ (and they would 
be significantly extincted in the mid-infrared if that were the case, leading
to their infrared luminosities being underestimated). Second, the majority of 
the obscured quasars possess luminous narrow emission lines, showing that 
a significant fraction of lines of sight from the nucleus are able to escape
to ionize the narrow-line gas. Furthermore, in the case
of radio-loud objects, where the radio lobe
luminosity provides an independent measure of the AGN luminosity, Haas
et al.\ (2004b) find that radio-loud quasars and their type-2 equivalents,
radio galaxies, show a statistically-identical ratio of radio to 10-1000$\mu$m
infrared luminosity, indicating that the obscured population has a similar
dust covering factor to the unobscured population, as expected from 
orientation-based Unified schemes (Antonucci 1993). We are thus 
confident that mid-infrared selection does not significantly bias an
AGN sample towards objects with unusually high dust covering factors, and 
so we are fairly confident
that our AGN sample is a fair proxy for one selected on the basis of \
bolometric flux. We can thus safely conclude that
$\sim 67$\% of black hole growth in the redshift and luminosity range
probed in this study is obscured by $E(B-V) \stackrel{>}{_{\sim}}0.3$.

\subsection{Comparison to other {\em Spitzer} selected samples}

Mid-infrared selection of obscured AGN 
has been successfully performed by several
other groups. Stern et al.\ (2005) use Hectospec data on the B\"{o}otes
field to study AGN selected using a mid-infrared color cut 
similar to that in Figure 1, but with different axes. Plotting the
objects of this paper onto the selection of Stern et al.\ 
we find that only a few obscured AGN found within our mid-infared color 
criteria are missed by their color selection.
They find an $\approx 3:1$ ratio of 
obscured to unobscured AGN, somewhat higher than found in this study, 
which is $\approx 2:1$. 
However, as Richards et al.\ (2006) point out, their result relies on 
a power-law interpolation of the quasar SED between the mid-infrared
and optical, which is not quite correct (see, e.g., the variation of
R-[8.0] color in Figure 5), so their 
obscured AGN fraction may be overestimated.
Alonso-Herrero et al.\ (2006) use another slightly different 
mid-infrared selection, based on the presence of a 
power-law mid-infrared SED, to find AGN in the Chandra Deep Field South. 
Due to this rather more complicated selection criterion it is hard to assess 
how many objects in the color-selected AGN samples would satisfy the
selection criteria of Alonso-Herrero et al., 
but most of them are likely to, as our color criterion is very 
similar to a power-law selection.
Their study complements the study of this paper very well, as their 24$\mu$m
fluxes range from $\approx 0.05-3$mJy compared to our range of 
$\approx 4.4-20$mJy.
Their median redshift of $\sim 1.4$ (based mostly on photometric redshifts), 
compares with a median of 0.6 for the combined spectroscopic
XFLS and SWIRE samples. Their objects are thus typically of similar, or 
only slightly lower luminosity to the objects in this study. 
The fact that they find a similar 
ratio of obscured to unobscured AGN of $\approx 2:1$ (based on SED fitting) 
is therefore consistent with no strong 
dependence on this ratio on redshift in the range $\sim 0.5-2$.

Mart\'\i nez-Sansigre et al.\ (2005,2006) use a 
different approach to selecting their obscured AGN. By searching for 
24$\mu$m sources with radio emission in excess of the radio-infrared 
correlation they are able to isolate obscured AGN irrespective of their 
mid-infrared SED shape. Although the properties of these radio-intermediate
objects may be different from the predominately radio-quiet populations 
found from the mid-infrared color selection, the ratio of obscured to 
unobscured objects is similar, in the range 1-3:1.

In the X-ray, Alonso-Herrero et al.\ (2006), Donley et al.\ 
(2005) and Polletta et al.\ (2006) 
find a significant population (40-60\%) of {\em Spitzer} selected AGN
which are highly-obscured in the X-ray. In 
particular, many {\em Spitzer}-selected AGN in the samples
of Alonso-Herrero et al.\
and Donley et al.\ are missing from even the $\sim $Ms
exposures of the {\em Chandra} Deep Fields. 
21 of the objects
in our samples overlap {\em XMM-Newton} fields, 
we detect 14 of them in $\sim 10-50$ks exposures. 
Given the difference in flux limits at 24$\mu$m, the relative X-ray depths 
are comparable, although the higher redshifts of the Alonso-Herrero
et al.\ and Donley et al.\ samples should lead to these objects being
more detectable at X-ray wavelengths due to the negative $k$-correction of
an absorbed X-ray spectrum. 

One final point of interest is that the three highest redshift objects in our
sample are all obscured quasars. Although the statistics are small, the 
fact that these objects were found in spite of strong selection effects
in favor of lower redshift objects is an indication that at very 
high redshifts ($z\stackrel{>}{_{\sim}}3$) obscured accretion may be even more
common.

\section{Conclusions}

The results presented in this paper, taken together with studies of higher redshift
objects selected from mid-infrared deep fields and objects found using joint
mid-infrared -- radio selection show that a large population of luminous 
obscured AGN exist at $z\stackrel{>}{_{\sim}}0.3$, including a significant population with 
bolometric luminosities above the Seyfert-quasar divide. We find that
the obscured population dominates in number density over the unobscured 
one over a wide range in AGN luminosity, from Seyferts to luminous quasars.
Comparison with Alonso-Herrero et al.\ (2006) suggests little variation in the 
obscured to unobscured ratio with redshift for AGN with strong mid-infrared
emission. 

Our results contrast with X-ray studies, which typically find that obscured AGN
are confined to low redshifts ($z\stackrel{<}{_{\sim}} 1$) and luminosities 
below the Seyfert-quasar divide 
$L_X \stackrel{<}{_{\sim}} 10^{44} {\rm ergs^{-1}}$ 
(e.g.\ Trump et al.\ 2006), suggesting that high luminosity and high redshift 
obscured quasars are missing from current X-ray based surveys, presumably due to 
limitations of areal coverage and depth. Although objects with columns 
$N_H \sim 10^{23}{\rm cm^2}$ can be found in deep X-ray surveys, locally $\sim 50$\%
of AGN are Compton thick, with columns $> N_H \sim 10^{24}{\rm cm^{-2}}$, and the equivalents of these
objects at high redshifts would be very hard to find in the X-ray, even allowing for the positive 
k-correction. This has important implications for models of the X-ray background
(e.g.\ Worseley et al.\ 2006).
However, it is  clear that all the techniques used to find obscured 
AGN are incomplete. Mid-infrared selection presumably fails for weak AGN, for highly 
obscured AGN, and at high 
redshifts. Our highest redshift objects are selected not on the basis of their
hot dust emission, but on the basis of dust-reddened accretion disk 
emission, so we 
will inevitably miss the most obscured objects at 
$z\stackrel{>}{_{\sim}}3$. 

For most objects, 
classification of galaxy type (AGN or starburst) based on optical emission 
line ratios and classification based on mid-infrared IRAC color appear 
to agree fairly well. The 
mid-infrared criterion is seen to fail for $z\stackrel{<}{_{\sim}} 0.3$
AGN with strong PAH 
emission in the IRAC 8.0$\mu$m band from an accompanying starburst. There
are also some objects with AGN-like IRAC colors which lack the high-ionization
emission line spectra of AGN. For most of these, their nature (AGN or 
starburst) remains to be resolved.

The discovery of a large population of obscured AGN has 
important implications for accretion efficiency and black hole demographics. 
The mass density of black holes in the Universe today may be accurately
estimated using the bulge-luminosity black hole mass relation, and the 
amount of accretion luminosity arising from unobscured accretion may be
estimated from the optical quasar luminosity function (Soltan 1982). 
Recent estimates using this technique place the radiative efficency of 
accretion close to that expected from a non-spinning (Schwartzschild)
black hole (e.g.\ Yu \& Tremaine 2002). 
However, if most accretion is actually hidden, to maintain the
same mass density of black holes today in the face of 
more accretion activity requires a {\em higher} radiative
efficiency, which can only be obtained from rapidly spinning 
Kerr black holes.

\acknowledgments

We would like to thank Alejo Martinez-Sansigre and Steve Rawlings for 
obtaining the WHT spectra, Michael Gregg for 
obtaining the Lick spectra, and the anonymous referee for a helpful 
report. The SWIRE team are thanked for producing 
and making available their {\em Spitzer} catalogs. 
Most of the optical data were 
obtained at the Hale Telescope, Palomar Observatory,
as part of a continuing collaboration between the California Institute of 
Technology (CIT), the Jet Propulsion Laboratory (JPL) 
(operated by CIT for the National Aeronautics and Space Administration 
[NASA]), and Cornell University. ML and AP were visiting astronomers
at the IRTF, which is operated by the University of Hawaii under Cooperative Agreement no. NCC 5-538 with the National Aeronautics and Space Administration, Science Mission Directorate, Planetary Astronomy Program.
This paper is based on observations made with the {\em Spitzer Space Telescope}
which is operated by JPL, CIT under a NASA contract. 
Support for this work was provided
by NASA through JPL. Funding for the creation and distribution of the SDSS 
Archive has been provided by the Alfred P. Sloan Foundation, the 
Participating Institutions, NASA, 
the National Science Foundation, the U.S. Department of 
Energy, the Japanese Monbukagakusho, and the Max Planck Society. 
This research has made extensive use of the NASA/IPAC Extragalactic 
Database (NED) which is operated by JPL, CIT, under contract with NASA.
Some of the data presented herein were obtained at the W.M. Keck Observatory, 
which is operated as a scientific partnership among CIT, 
the University of California and NASA. 
The authors wish to recognize and acknowledge the very significant cultural 
role and reverence that the summit of Mauna Kea has always had within the 
indigenous Hawaiian community.  

\appendix

\section{Comparison of classifications with those of Papovich et al.\ }

In an independent study, Papovich et al.\ (2006) obtained spectra
of bright 24$\mu$m XFLS sources with the Hectospec Fibre spectrograph on the 
MMT. Several of these objects are common to our XFLS AGN sample. 
Papovich et al.\ attempted to classify their objects using a semi-automated
algorithm. For the most part, our classifications agree, and where
we differ it is usually because the Papovich et al.\ ``galaxy'' 
classification generally includes narrow-line AGN (Table 6). 
There are, however, a few objects classed as broad line 
quasars or AGN for which there seems to be no evidence for broad 
lines in either our spectra or the Hectospec spectra. In one case, however, 
(SSTXFLS J171430.7+584225) the Hectospec spectrum does show weak broad H$\beta$
and Mg{\sc ii}2798 emission, which appear very weak or absent in our 
Palomar/COSMIC spectrum. The Hectospec spectrum is of higher signal-to-noise
than our spectrum, and the H$\beta$ line falls close to the 7600\AA$\;$
atmospheric absorption feature (for which we do not attempt to
correct), so it is most likely that we simply missed the weak broad H$\beta$
line in this case. We have therefore reclassified this object as type-1R. 
One further discrepancy is in our classification of SSTSFLS J171454.4+584948,
ths object is classed at QSO/starburst by Papovich et al., but plots among
the starbursts the BPT plot of Figure 4b, and has thus been classified as a 
starburst.

\section{Spectroscopy of the 8$\mu$m flux limited sample}

Most of the objects in the 8$\mu$m flux-limited sample
of candidate obscured AGN of Lacy et al.\ 
(2004, 2005a) are also members of the XFLS sample of Table 3, however, 
a few have 24$\mu$m fluxes below 4.6mJy and are hence missing. For completeness
we list their redshifts, where measured, 
and spectral types in Table 7 (see also Lacy et al.\ 
2005).

\begin{table*}
{\scriptsize
\caption{Non-SDSS objects in the XFLS sample with spectra in 
Papovich et al.\ (2006)}
\begin{tabular}{lccccll}
Name &$S_{24.0}{\rm /mJy}^{1}$ & Redshift&Class&Basis & P06 class$^{*}$\\ \hline
SSTXFLS J171115.2+594906 & 9.4 &0.587&Sbt  & & galaxy\\
SSTXFLS J171147.4+585839 & 4.8 &0.800&2  &[Ne{\sc v}]13&galaxy\\
SSTXLFS J171324.2+585549 & 4.9 &0.609&2  &[Ne{\sc v}]12&galaxy\\
SSTXFLS J171325.1+590531 & 9.5 &0.126&2  &BPT&QSO/AGN\\
SSTXFLS J171331.5+585804 & 5.9 &0.435&Sbt/L  &&Galaxy\\
                         &     &     &   &            & \\
SSTXFLS J171335.1+584756 &23.7 &0.133&1R&Fe{\sc ii}&QSO/BL\\
SSTXFLS J171430.7+584225 & 8.2 &0.561&2  &[Ne{\sc v}]20&QSO/BL\\
SSTSFLS J171454.4+584948 & 4.9 &0.253&Sbt &BPT& Galaxy/starforming\\
SSTXLFS J171513.8+594638 & 5.0 &0.248&1R &BL (radio-loud)&QSO/starburst\\
SSTXFLS J171530.7+600216 &11.6 &0.420&2 &[Ne{\sc v}]26&QSO/BL\\
                         &     &     &   &            & \\
SSTXFLS J171650.6+595752 & 6.7 &0.182&Sbt  &BPT&galaxy/starburst\\
SSTXFLS J171708.6+591341 & 5.4 &0.646&2  &[Ne{\sc v}]37&QSO/BL\\
SSTXFLS J171839.6+593359 &10.9 &0.383&1 &BL (radio loud)&QSO/BL\\
SSTXFLS J171913.5+584508 & 8.9 &0.318&2  &[Ne{\sc v}]12, BPT&QSO\\
SSTXFLS J172219.5+594506 & 7.8 &0.271&2  &BPT&QSO/AGN\_BROAD\\
                         &     &      &   &   &   \\
SSTXFLS J172228.1+601526 & 7.1 &0.741&2  &[Ne{\sc v}]&Galaxy/manual \\
SSTXFLS J172245.0+590328 & 4.7 &0.797&2  &[Ne{\sc v}]43&Galaxy\\
\end{tabular}

\noindent
$^{*}$ P06 class is the classification of the object in the MMT/Hectospec survey of Papovich et al.\ (2006), in the form class/subclass.
}

\end{table*}

\begin{table*}
\caption{Objects in the 8$\mu$m selected sample of Lacy et al.\ 2004 not 
in the 24$\mu$m flux-limited XFLS sample.}
{\scriptsize
\begin{tabular}{lcccc}
Name                  &$S_{8.0}$&$z$ &Class&Basis\\
                      &(mJy)    &    &     &\\\hline 
SSTXFLS J171106.8+590436&1.38&0.462   & 2?  &[O{\sc iii}]5007/[O{\sc ii}]3727=3.8\\
SSTXFLS J171133.4+594906&1.91&  ?     & 1?  & -\\
SSTXFLS J171421.3+602239&1.25&  ?     & 2/Sbt/L?  &\\
SSTXFLS J171804.6+602705&1.05& 0.43?  & 2/Sbt/L?  &\\
SSTXFLS J171930.9+594751&1.44& 0.358  & 2& [Ne{\sc v}]30\\ 
SSTXFLS J172253.9+582955&1.00& 1.753$^{*}$  & 1    &BL\\
SSTXFLS J172458.3+591545&1.10& 0.494  & 2?   &[O{\sc iii}]5007/[O{\sc ii}]3727=1.8\\
SSTXFLS J172601.8+601100&1.43& 1.12?  & 1R   & single BL\\       
\end{tabular}

\noindent
$^{*}$ spectrum from Papovich et al.\ (2006).
}
\end{table*}




\end{document}